\documentclass[prd,aps,10pt,superscriptaddress,twocolumn,amsmath,amssymb,nofootinbib,longbibliography, floatfix]{revtex4-1}

\usepackage{ragged2e}

\usepackage[english]{babel}

\usepackage{natbib}
\usepackage[colorlinks]{hyperref}
\hypersetup{linkcolor=blue,citecolor=blue,filecolor=blue,urlcolor=blue}

\usepackage{amssymb,amsmath,multirow}
\usepackage{bm}
\usepackage{dsfont}
\usepackage{orcidlink}
\usepackage{color}

\usepackage[utf8]{inputenc}
\usepackage{times}
\usepackage{setspace}
\usepackage{floatrow}
\usepackage{caption}
\usepackage{booktabs}
\usepackage{multirow}
\usepackage{siunitx}

\usepackage[normalem]{ulem}
\usepackage{comment}

\usepackage{subfig,hyperref}

\usepackage{graphicx}
\usepackage{dcolumn}
\usepackage{bm}
\usepackage{wasysym}
\usepackage[
   starfontserif 
    ]{starfont}
\usepackage{amsmath}

\DeclareSymbolFont{starfontsym}{OT1}{sts}{m}{n}
\DeclareMathSymbol{\mathTerra}{\mathord}{starfontsym}{76}

\newcommand{\beq}{\begin{equation}}
\newcommand{\beqn}{\begin{align}}
\newcommand{\eeq}{\end{equation}}
\newcommand{\eeqn}{\end{align}}

\begin{document}

\title{Hawking Radiation Signatures from Primordial Black Holes Transiting the Inner Solar System: Prospects for Detection}

\author{Alexandra P. Klipfel \orcidlink{0000-0002-1907-7468}}
 \email{aklipfel@mit.edu}
\affiliation{%
Department of Physics, Massachusetts Institute of Technology, Cambridge, MA 02139, USA}
 
\author{Peter Fisher}
 \email{fisherp@mit.edu}
\affiliation{%
Department of Physics, Massachusetts Institute of Technology, Cambridge, MA 02139, USA}

\author{David I.~Kaiser \orcidlink{0000-0002-5054-6744}}
 \email{dikaiser@mit.edu}

\affiliation{%
Department of Physics, Massachusetts Institute of Technology, Cambridge, MA 02139, USA}

\date{\today}

\begin{abstract}
Primordial black holes (PBHs) arise from the collapse of density perturbations in the early universe and serve as a dark matter (DM) candidate and a probe of fundamental physics. There remains an unconstrained ``asteroid-mass'' window where PBHs of masses $10^{17} {\rm g} \lesssim M \lesssim 10^{23} {\rm g}$ could comprise up to $100\%$ of the dark matter. Current $e^{\pm}$ Hawking radiation constraints on the DM fraction of PBHs are set by comparing observed spatial- and time-integrated cosmic ray flux measurements with predicted Hawking emission fluxes from the galactic DM halo. These constraints depend on cosmic ray production and propagation models, the galactic DM density distribution, and the PBH mass function. We propose to mitigate these model dependencies by developing a new local, time-dependent Hawking radiation signature to detect low-mass PBHs transiting through the inner Solar System.  We calculate transit rates for PBHs that form with initial masses $M \lesssim  5\times10^{17}\text{g}$. We then simulate time-dependent positron signals from individual PBH flybys as measured by the Alpha Magnetic Spectrometer (AMS) experiment in low-Earth orbit. We find that AMS is sensitive to PBHs with masses $M\lesssim 2\times10^{14} \, {\rm g}$ due to its lower energy threshold of $500 \, {\rm MeV}$. We demonstrate that a dataset of daily positron fluxes over the energy range $5-500 \, {\rm MeV}$, with similar levels of precision to the existing AMS data, would enable detection of PBHs drawn from present-day distributions that peak within the asteroid-mass window. Our simulations yield ${\cal O} (1)$ detectable PBH transits per year across wide regions of parameter space, which may be used to constrain PBH mass functions. This technique could be extended to detect $\gamma$-ray and X-ray Hawking emission to probe further into the asteroid-mass window.
\end{abstract}

\maketitle

\section{\label{sec:Introduction}Introduction}

Primordial black holes (PBH) are theorized to form from the collapse of primordial density perturbations in the early universe. Those that form with sufficient initial mass, $M_i \gtrsim 5.364\times10^{14} \, {\rm g}$, would persist to the present day and contribute to the dark matter (DM) density $\Omega_{\rm DM}$  \cite{zeldovich_hypothesis_1966,hawking_gravitationally_1971,carr_black_1974,carr_cosmological_1976,Khlopov:1985fch,Khlopov:2008qy,escriva_primordial_2024}. Various cosmological models predict the existence of PBHs over a wide present-day mass range which, in principle, could span approximately $40$ orders of magnitude, from $10^{14} \, {\rm g}$ to $10^{55} \, {\rm g}$ \cite{Carr:2009jm,carr_primordial_2020,carr_constraints_2021,carr_primordial_2022,escriva_primordial_2024,Ozsoy:2023ryl}. This translates to $10^{-19}M_{\odot} \lesssim M \lesssim 10^{22}M_{\odot}$.  Despite many experimental constraints on the PBH dark matter fraction $f_{\rm PBH}$, there remains an open ``asteroid-mass'' window within which PBHs of mass $10^{17} \, {\rm g} \lesssim M \lesssim 10^{23} \, {\rm g}$ may comprise up to 100\% of the cold dark matter (CDM) \cite{Carr:2009jm,carr_primordial_2020,carr_constraints_2021,carr_primordial_2022,green_primordial_2021,carr_observational_2024,gorton_how_2024,escriva_primordial_2024,Ozsoy:2023ryl,Boybeyi:2024mhp}. We are motivated to experimentally search for PBHs not only because they could make up a significant fraction of the dark matter, but because the presence or absence of PBHs in our universe, even for $f_{\rm PBH} \ll 1$, can allow us to constrain cosmological models and to probe fundamental and beyond-Standard-Model (BSM) physics. 

Studying PBHs within the framework of individual transit events through our Solar System provides unique opportunities to constrain the PBH dark matter fraction, directly observe low-mass PBHs, and detect Hawking radiation. In this article we investigate the feasibility of using Hawking emission from PBHs transiting through the inner Solar System to constrain the PBH dark matter fraction for $M\lesssim 5 \times 10^{17}$ g.  We investigate whether a time-dependent positron emission signature from a PBH transiting within approximately $10^{-3}-10 \, {\rm AU}$ of the Earth would be detectable with the Alpha Magnetic Spectrometer (AMS) experiment, a cosmic ray observatory mounted on the International Space Station (ISS) \cite{aguilar_alpha_2021}, or with realistic next-generation detectors.  We also discuss the possibility of performing similar analyses with high-energy photons and $e^{\pm}$ datasets from various other telescopes, cosmic ray observatories, and solar probes. This new time-dependent PBH transit signature is valuable, because it is model-independent and could be extended to X-ray and low-energy positron datasets to probe far into the asteroid-mass window.

Upon developing a matched-filter pipeline to extract time-dependent Hawking emission signals from simulated time-series AMS positron data, we find that the existing AMS experiment is only sensitive to transiting PBHs with masses $M \lesssim 2\times10^{14} \, {\rm g}$. As discussed below, such a value is well below the present lower-bound for which PBHs could constitute a significant fraction of the DM abundance today. The positron energy range to which the AMS experiment is sensitive ($Q \geq Q_{\rm min} = 500 \, {\rm MeV}$) is limited by the strong effects of the geomagnetic field for detectors (like AMS) in low-Earth orbit. We show that realistic detectors with AMS-like precision positioned at higher altitudes, or even at a Lagrange point, 
could reliably detect positrons with $Q \geq Q_{\rm min} = 5 \, {\rm MeV}$. Such instruments would be capable of directly detecting several PBH transits per year, or---in the absence of any such detections---of placing strong constraints on a large set of PBH mass functions.

Following a brief discussion of the motivation and current constraints on searches for Hawking radiation from PBHs, in Sec.~\ref{sec:Models} we introduce various models and parameterizations for Hawking emission, PBH mass distributions, the local DM density distribution, and detector characteristics. Section~\ref{sec:Results} discusses expected transit rates for positron-producing PBHs in the inner Solar System, simulation results for time-dependent positron Hawking emission signatures from PBH transits, and prospects for detecting PBHs and/or constraining PBH mass functions with existing and hypothetical future datasets. Concluding remarks follow in Sec.~\ref{sec:Conclusion}.

\subsection{\label{sec:Motivation}Motivation}

Primordial black holes garnered new attention as a dark matter candidate over the last decade in the wake of LIGO-Virgo-KAGRA (LVK) collaboration data and James Webb Space Telescope (JWST) observations. Some physicists \cite{clesse_gw190425_2021, huang_gw230529_181500_2024} have interpreted gravitational-wave signatures from binary black hole mergers with candidates in the lower \cite{the_ligo_scientific_collaboration_gwtc-3_2023} and pair-instability \cite{the_ligo_scientific_collaboration_gw190521_2020} mass gaps as potential PBH candidates.  Additionally, several analyses of public LVK data claim to observe sub-solar mass black holes \cite{phukon_hunt_2021, morras_analysis_2023}, which would be excellent PBH candidates. The LVK collaboration, however, reports no such candidates with sufficient significance in their second and third observing runs \cite{the_lvk_collaboration_search_2023, ligo_scientific_collaboration_and_virgo_collaboration_search_2018}. Furthermore, JWST observations of galaxies \cite{castellano_early_2022, castellano_early_2023, adams_discovery_2022, atek_revealing_2022, harikane_comprehensive_2023, naidu_two_2022} and candidate active galactic nuclei (AGN) \cite{bogdan_evidence_2023, natarajan_first_2023, kovacs_candidate_2024, larson_ceers_2023, maiolino_small_2024} with redshift $z \gtrsim 10$ are in tension with the $\Lambda$CDM standard model of cosmology \cite{menci_high-redshift_2022}. These observations have reinvigorated interest in PBHs as seeds for early galaxies and the supermassive black holes at the centers of their AGNs \cite{dolgov_solution_2024, hutsi_did_2023, silk_which_2024, liu_accelerating_2022}.  

The $\Lambda$CDM model predicts that about $26\%$ of the energy density of the universe is comprised of 
non-relativistic, non-baryonic matter with extremely weak coupling to Standard Model particles \cite{planck_collaboration_planck_2020}. Whereas the evidence for the existence and relative fraction of DM is very strong, as yet there does not exist any decisive evidence to support various DM particle candidates, such as axions, axion-like particles, weakly interacting massive particles (WIMPs), or sterile neutrinos \cite{Boveia:2018yeb,Weltman:2018zrl,Agrawal:2021dbo,Cooley:2022ufh,ParticleDataGroup:2022pth}. Primordial black holes also satisfy the $\Lambda$CDM dark matter criteria, and their formation and properties can be explained entirely within current theories of general relativity, $\Lambda\text{CDM}$, inflation, and the Standard Model. Most models predict that PBHs formed before baryogenesis and thus contribute to the non-baryonic dark matter density $\Omega_{\text{DM}}$ rather than to the baryonic density $\Omega_b$---which can be at most $5\%$ of the critical energy density \cite{carr_primordial_2020}.  Because PBHs with masses within the asteroid-mass range (or larger) are expected to be 
uncharged under any Standard Model gauge group in the present-day epoch \cite{carter_charge_1974, alonso-monsalve_primordial_2024}, they can only interact gravitationally with SM particles. Furthermore, due to their large masses, PBHs are inherently non-relativistic---``cold''---objects capable of coalescing to form large-scale structures in the universe.

Primordial black holes also provide a useful tool with which to investigate cosmological history. PBHs would form from the collapse of primordial overdensities. Several mechanisms have been proposed that could amplify primordial overdensities on appropriate length-scales to seed PBH formation, including a sudden change in the effective sound speed at the time of the QCD phase transition, domain-wall collapse, self-intersection of cosmic strings, or bubble collisions from strongly first-order phase transitions \cite{JedamzikPBHpto,LiuPBHpto,Baker:2021nyl,Baker:2025cff,Lewicki:2023ioy,Goncalves:2024vkj}. (See Refs.~\cite{carr_primordial_2022, green_primordial_2021,escriva_primordial_2024,Ozsoy:2023ryl} for a thorough discussion of possible PBH formation processes.) 

Among the best-studied formation mechanisms involve density perturbations that were amplified during a phase of early-universe inflation. When overdensities above some critical threshold re-enter the Hubble radius following the end of inflation, they induce gravitational collapse into black holes. In these scenarios, the overdensities are associated with a peak in the spectrum of primordial gauge-invariant curvature perturbations \cite{Niemeyer:1997mt,Green:1999xm,Young:2019yug,Escriva:2019phb,Gow:2020bzo,Musco:2020jjb}. (See especially the reviews in Refs.~\cite{Gundlach:2007gc,Escriva:2021aeh}.) Such peaks can form from dynamics during inflation that depart from ordinary slow-roll evolution, such as a transient phase of ultra-slow-roll dynamics \cite{Kinney:2005vj,garcia-bellido_primordial_2017,Motohashi:2017kbs,Di:2017ndc,Pattison:2021oen,Geller:2022nkr} or brief tachyonic growth of perturbations on short length-scales associated with rapid turns in field space (within multifield models) \cite{garcia-bellido_density_1996,Lyth:2010zq,Halpern:2014mca,clesse_massive_2015,Fumagalli:2020adf,Braglia:2020eai,Palma:2020ejf,Lorenzoni:2025gni}. Recent works have demonstrated that such dynamics can arise within models of the early universe that incorporate realistic features from high-energy physics, including multiple interacting fields, with less fine-tuning of parameters than typically required in simpler single-field constructions \cite{Qin:2023lgo,Lorenzoni:2025gni}.

The parameters of a specific cosmological model determine the shape of the present-day PBH number density distribution, $dn/dM$, and the fraction of DM comprised of PBHs, $f_{\rm PBH} = \Omega_{\rm PBH,0} / \Omega_{\rm DM,0}$. Working backward from experimental constraints on $f_{\rm PBH}$, we can constrain cosmological models and probe the physics of the early universe and the shape of the primordial power spectrum.

Even if $f_{\rm PBH} \ll 1$, primordial black holes can serve as a useful probe of BSM physics. Studies of mixed dark matter scenarios, in which PBHs and one or more additional candidates comprise all the dark matter, place strong---nearly mutually exclusive---constraints on WIMPs in the GeV-TeV mass range for a wide range of PBH masses and DM fractions \cite{carr_primordial_2020}. Mixed DM scenarios with axion-like particles \cite{kuhnel_decaying_2019} or sterile neutrinos \cite{kuhnel_signatures_2017} may also admit mutual constraints on these particle DM candidates and PBHs.

Perhaps most importantly, observation of low-mass PBHs may be the only viable method to directly detect Hawking radiation \cite{hawking_black_1974,hawking_particle_1975}, which is widely accepted as a theoretical triumph due to its unification of statistical mechanics, quantum field theory, and general relativity into a semi-classical framework. Black holes with masses larger than that of the Earth, $M \gtrsim M_{\mathTerra} \approx 10^{27} \, {\rm g},$ emit only photons and neutrinos with temperatures significantly below $2.7 \, {\rm K}$ and $1.9 \, {\rm K}$, respectively. Therefore, Hawking radiation from all known stellar and supermassive black holes will be hidden by the Cosmic Microwave Background (CMB) and Cosmic Neutrino Background (C$\nu$B). However, the existence of PBHs with masses $M\ll M_{\mathTerra}$---even if $f_{\rm PBH} \ll 1$---would provide an opportunity to directly detect Hawking emission of photons, neutrinos, and $e^{\pm}$. Recent work also posits that extremely low-mass PBHs may emit heavier stable particles at detectable rates, including isotopes of hydrogen and helium nuclei \cite{fisher_primal_2025}.  Detecting Hawking radiation is among the only experimental probes with which physicists can investigate a possible unification of quantum field theory and general relativity at achievable energy scales.

\subsection{\label{sec:ExptlConstraints}Evaporation Constraints}

Across their entire mass range, PBHs should interact with known matter in predictable ways. These interactions manifest as observational signatures, which can be used to directly detect PBHs or to place constraints on $f_{\rm PBH}$. Potential signatures include Hawking emission, accretion, dynamical effects, large-scale structure formation, gravitational lensing, and gravitational waves. See Refs.~\cite{escriva_primordial_2024,Carr:2009jm, carr_primordial_2020,green_primordial_2021,carr_constraints_2021,carr_primordial_2022,Ozsoy:2023ryl,carr_observational_2024,gorton_how_2024,Boybeyi:2024mhp} for a thorough review of current constraints from these signatures and others. We are interested in developing a new PBH detection signature and the requisite data analysis techniques to probe the open asteroid-mass window, $10^{17} \text{ g} \leq M \leq 10^{23}$ g. Reaching this notoriously inaccessible PBH mass range requires the development of novel signatures because these black holes have Schwarzschild radii, $R_s \equiv 2GM/c^2$, smaller than a micron, thus rendering known gravitational lensing techniques to be much more difficult to apply.

The black holes we discuss in this article have initial formation-time masses $5.364\times 10^{14}\text{ g} \leq M_i \leq 5\times 10^{17}$ g, a window we refer to as the ``$e^{\pm}$-production mass range,'' which corresponds to radii between approximately $1-1000 \, {\rm fm}$. We are focusing our initial efforts on positron Hawking emission, and therefore on PBHs with present-day masses $M \leq 5\times 10^{17} \, {\rm g}$, because positrons have low astrophysical backgrounds and the AMS-02 experiment has a large acceptance and provides over a decade of high-precision time-series positron flux data. 


Hawking radiation provides useful \textit{evaporation constraints} on the dark matter fraction of low-mass PBHs with $M \lesssim 10^{17} {\rm g}$. Hawking's semi-classical formulation of black hole evaporation \cite{hawking_particle_1975,hawking_black_1974} predicts that a PBH should emit all SM particles during the course of its lifetime. PBHs that form within the $e^{\pm}$-producing range, $5.364\times10^{14} \leq M_i \leq 5\times10^{17}$ g, will emit photons, neutrinos \cite{klipfel_ultra-high-energy_2025,Lysyy:2024qrs}, light charged leptons, and pions peaking in the MeV-GeV energy range, while PBHs in the asteroid-mass window will emit photons and neutrinos peaking in the eV-MeV range. PBHs at the very end of their lives will rapidly lose their remaining mass and effectively explode by emitting high fluxes of all 17 Standard Model particles---including heavy hadrons, leptons, and massive bosons. For example, as discussed in section \ref{sec:PBHMassLoss}, a PBH with mass $M_* \equiv 5.364\times10^{14}$ g has a lifetime equal to the current age of the universe, while a PBH with mass $6\times 10^{10}$ g has a lifetime of approximately 1 day and a temperature $T \approx 176 \, {\rm GeV}$, which is hot enough to emit top quarks and Higgs bosons. PBH explosions could also be sources of ultra-high-energy cosmic rays \cite{Boluna:2023jlo,Baker:2025zxm,klipfel_ultra-high-energy_2025,Baker:2025cff,Airoldi:2025opo} and heavy antimatter cosmic rays \cite{fisher_primal_2025}.

Evaporation constraints in the literature from present-epoch Hawking radiation are split into two categories: $\gamma$-ray constraints and $e^{\pm}$ constraints. 
The methodology behind these constraints is to exclude all PBH mass-functions for which a galactic PBH population would produce particle fluxes that exceed the observed galactic (and extragalactic) time- and spatially-integrated cosmic ray backgrounds. Current $e^{\pm}$ Hawking emission constraints are derived from time-independent Voyager-1 data and Alpha Magnetic Spectrometer (AMS) data \cite{boudaud_voyager_2019, huang_constraints_2024, su_constraining_2024, luque_refining_2024}. Due to the uncertainties associated with galactic cosmic ray propagation models and our current limited understanding of the $e^{+}$ background, we seek to develop a new, local signature of PBH Hawking emission that does not rely so heavily on these models.

Searching for PBHs that transit through our Solar System allows us to be agnostic about PBH mass functions and galactic cosmic ray propagation models and affords an opportunity to make a direct detection of Hawking radiation. Previous literature proposes that detecting Hawking  radiation \cite{sobrinho_direct_2014, arbey_detecting_2021} or orbital perturbations \cite{tran_close_2023,Cuadrat-Grzybowski:2024uph,Thoss:2024dkg} from PBHs in our Solar System may be viable detection strategies. In this article, we simulate the observed time-dependent positron fluxes from PBHs transiting between $\sim0.01-10 \, {\rm AU}$
of the Earth and investigate whether such transits could be detected by the AMS-02 experiment aboard the ISS. We predict the PBH mass functions that could be constrained with current AMS data and with possible future cosmic ray positron datasets. Furthermore, the method of detecting time-dependent Hawking radiation signatures from local transits developed here can be generalized to $\gamma$-ray, X-ray, and low-energy positron datasets, which would allow us to probe PBHs with masses up to $\sim 10^{20}\, {\rm g}$---a mass range unreachable by current evaporation constraint techniques based on isotropic galactic and extragalactic cosmic ray fluxes.

Orbital perturbation signatures are most sensitive to the heaviest PBHs in the asteroid-mass range, while the proposed Hawking emission signatures discussed here are most sensitive to the lighter end of this window, where the PBHs are hotter emitters. Thus, these two novel signatures combined have the potential to probe the entirety of the asteroid-mass window, and to possibly make a direct detection of a PBH candidate or to place strong constraints on the PBH DM fraction. 

\section{\label{sec:Models}Models}
\subsection{\label{sec:HR}Hawking Radiation}

A black hole can couple to every form of matter because of the universal nature of gravitation.
Therefore, Hawking radiation \cite{hawking_black_1974,hawking_particle_1975,page_particle_1976,page_particle_1976-1,page_particle_1977} is not limited to the emission of photons. Provided that a black hole is small enough and therefore hot enough, it can emit all 17 Standard Model particles as fundamental degrees of freedom. These fundamental emission rates are referred to as \textit{primary emission spectra}. To determine the feasibility of detecting Hawking radiation from primordial black holes transiting past Earth with space-based cosmic ray experiments like AMS, we must accurately simulate the PBH emission spectra for photons, electrons, and positrons. Because our technique probes hot, low-mass PBHs, it is critical that we account for QCD jets, hadronization, and particle decays associated with the primary emission of quarks, muons and other short-lived particles \cite{macgibbon_quark-_1990}. Jet fragmentation and particle decays will contribute to the net emission rates---referred to as \textit{secondary emission spectra}---of stable, detectable fundamental and composite particles. We therefore will use secondary rather than primary emission spectra throughout our analysis. 

\subsubsection{\label{sec:PrimarySpectra}Primary Emission}

A black hole of mass $M$, angular velocity $\Omega$, electric potential $\Phi$, and surface gravity $\kappa$ will emit fundamental particles of species $j$ at a rate \cite{page_particle_1976, macgibbon_quark-_1990}:
\begin{equation}
    \label{eqn:PrimarySpectra}
    \frac{d^2N_j^{(1)}}{dtdQ} = g_j\frac{\Gamma_{s_j}}{2 \pi \hbar} \left[\exp{\left( \frac{Q - n \hbar \Omega - q \Phi}{\hbar \kappa/2 \pi c} \right)} - (-1)^{2s_j}  \right]^{-1}
\end{equation}
where the particle has spin $s$, angular momentum $n\hbar$, charge $q$, energy $Q$, and total quantum degrees of freedom $g$. The quantity in Eq.~(\ref{eqn:PrimarySpectra}) is called the \textit{primary emission spectrum} for a given particle species. (See the Supplemental Materials of Ref. \cite{klipfel_ultra-high-energy_2025} for appropriate $g$ parameters for all SM particles.) To compute the primary emission spectrum of positrons, for example, we must use $g=2$ to account for the two helicity degrees of freedom. 
The bracketed term on the right hand side of Eq.~(\ref{eqn:PrimarySpectra}) has the form of a blackbody spectrum for an emitter at temperature
\begin{equation}
    \label{eqn:BHTempGen}
     T = \frac{\hbar \kappa}{2 \pi k_B c}. 
\end{equation}

The important physics enters through the \textit{greybody factor}, $\Gamma_s$, a dimensionless quantity that governs the interaction between the black hole and the emitted particles, which is generally a function of the black hole parameters $\Omega, \ \Phi, \ M,$ and $\kappa$ and the particle parameters $Q, \ m,$ and $ s$. The greybody factor is defined in terms of the absorption cross section $\sigma_s$ associated with scattering a
Standard Model field of mass $m$ and spin $s$ off a black hole of mass $M$ in curved spacetime \cite{macgibbon_quark-_1990}: 
\begin{equation}
    \label{eqn:GreyBody}
    \Gamma_s(M, Q, m) = \frac{\sigma_s(M, Q, m)}{\pi \hbar^2 c^2}(Q^2 - m^2).
\end{equation}
A simplified form of Eq.~(\ref{eqn:GreyBody}) holds exactly for massless particles and approximately for relativistic massive particles: 
\begin{equation}
    \label{eqn:GreyBody2}
    \Gamma_s(M, Q) = \frac{\sigma_s(M, Q)}{\pi \hbar^2 c^2}Q^2.
\end{equation}
(For a perfect blackbody, $\Gamma_s = 1.)$ Note that under the relativistic assumption of Eq.~(\ref{eqn:GreyBody2}), the emission spectra of massive particles must be cut off at $Q=m$. MacGibbon and Webber \cite{macgibbon_quark-_1990} remark that the relativistic form of $\Gamma_s$ in Eq.~(\ref{eqn:GreyBody2}) only differs notably from the exact non-relativistic result for $Q \leq 2m$, and never by more than 50\%. We thus conclude that the cutoff at $Q=m$ dominates calculations of total emission rates and we can employ the relativistic approximation of $\Gamma_s$ for all $Q > m$. The calculation of greybody factors for Schwarzschild black holes is discussed in detail in Refs.~\cite{page_particle_1976, macgibbon_quark-_1990, teukolsky_perturbations_1974, teukolsky_perturbations_1973}. 


Because we only consider non-rotating and uncharged Schwarzschild black holes, we take $\Omega=\Phi=0$ and 
\begin{equation}
    \label{eqn:BHTemp}
    k_B T = \frac{\hbar c^3}{8 \pi G M}.
\end{equation}
Assuming also that $\Gamma_s$ takes the form of Eq.~(\ref{eqn:GreyBody2}), the primary emission spectrum for particle species $j$ reduces to:
\begin{equation}
    \label{eqn:PrimarySpectraSchwarzschild}
    \frac{d^2N_j^{(1)}}{dtdQ} = g_j \frac{\Gamma_{s_j}}{2 \pi \hbar} \left[\exp{\left( \frac{Q}{\hbar c^3/8\pi G M} \right)} - (-1)^{2s_j}  \right]^{-1}.
\end{equation}

The assumption that primordial black holes are uncharged and non-rotating is a reasonable one. Page \cite{page_particle_1976-1} showed that an initially rotating black hole will preferentially emit particles with spin aligned with the hole's angular momentum, causing it to ``spin down'' to within $1\%$ of the emission rate a Schwarzschild configuration before half of its energy is radiated away. He further demonstrates that an initially maximally rotating PBH will only have a lifetime $2.32$ times longer than its Schwarzschild counterpart \cite{page_particle_1976-1}. Page thus concludes that it is valid to assume decaying black holes have negligible rotation \cite{page_particle_1976-1}. In addition, PBHs generally form with $\Omega=0$ from the collapse of spherically symmetric scalar perturbations, and, due to negligible accretion rates, PBHs within the asteroid-mass range 
will not spin up over time \cite{chiba_spin_2017, de_luca_evolution_2020, chongchitnan_extreme-value_2021}.  Furthermore, PBHs discharge any initial charge even more efficiently than they lose angular momentum \cite{gibbons_vacuum_1975,carter_charge_1974}, leaving tiny, stochastic charge fluctuations of order ${\cal O} (e) \ll \sqrt{G}\, M$ \cite{page_particle_1977,vasquez_corrections_2024}.

We therefore assume all PBHs are Schwarzschild and use Eq.~(\ref{eqn:PrimarySpectraSchwarzschild}) to compute primary emission spectra for all 17 fundamental Standard Model particles. The calculation of primary spectra is done numerically via the open source Hawking radiation code \texttt{BlackHawk v2.2}, which makes equivalent assumptions \cite{arbey_blackhawk_2019, arbey_physics_2021}.

\begin{figure}[]
    \centering
    \includegraphics[width=0.95\textwidth]{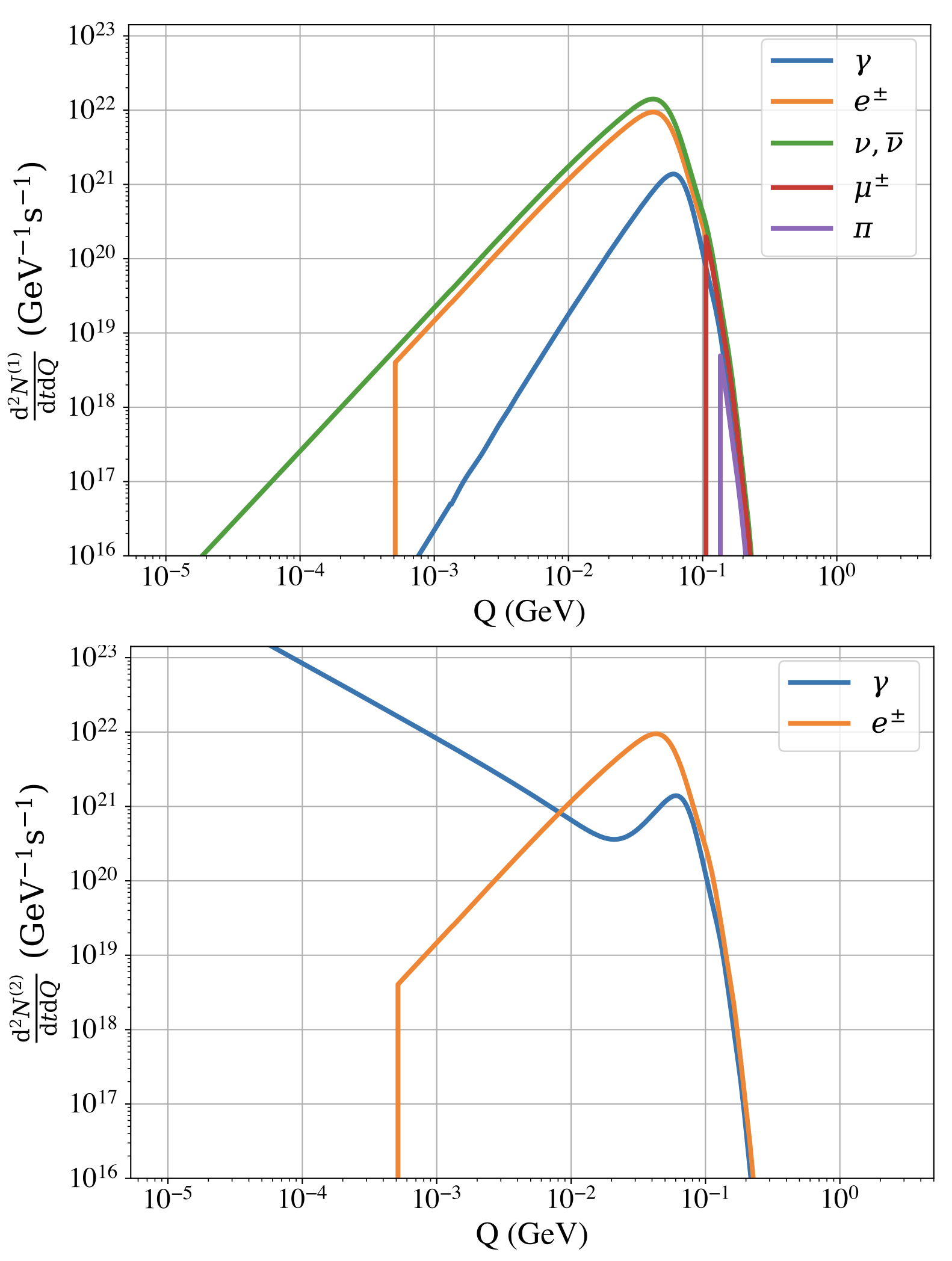}
    \caption{\justifying Primary (top) and secondary (bottom) Hawking emission spectra for a primordial black hole with mass $M=10^{15}$g. Spectra are calculated numerically with \texttt{BlackHawk v2.2} \cite{arbey_blackhawk_2019, arbey_blackhawk_nodate} and \texttt{hazma} \cite{coogan_hazma_2020} following the discussions in Sections \ref{sec:PrimarySpectra} and \ref{sec:SecondarySpectra}. Note that no secondary $p$ or $\bar{p}$ appear because these rates are negligible given $T = 10.6 \text{ MeV} \ll m_p$. Fermion spectra shown are the sum of particle and antiparticle emission rates. The primary neutrino spectra are summed over all flavors. No secondary neutrino spectra are shown because such processes are not incorporated within the  \texttt{hazma} code.}
    \label{fig:SpectraComparison}
\end{figure}

\begin{figure}[]
\centering
\includegraphics[width=1.0\textwidth]{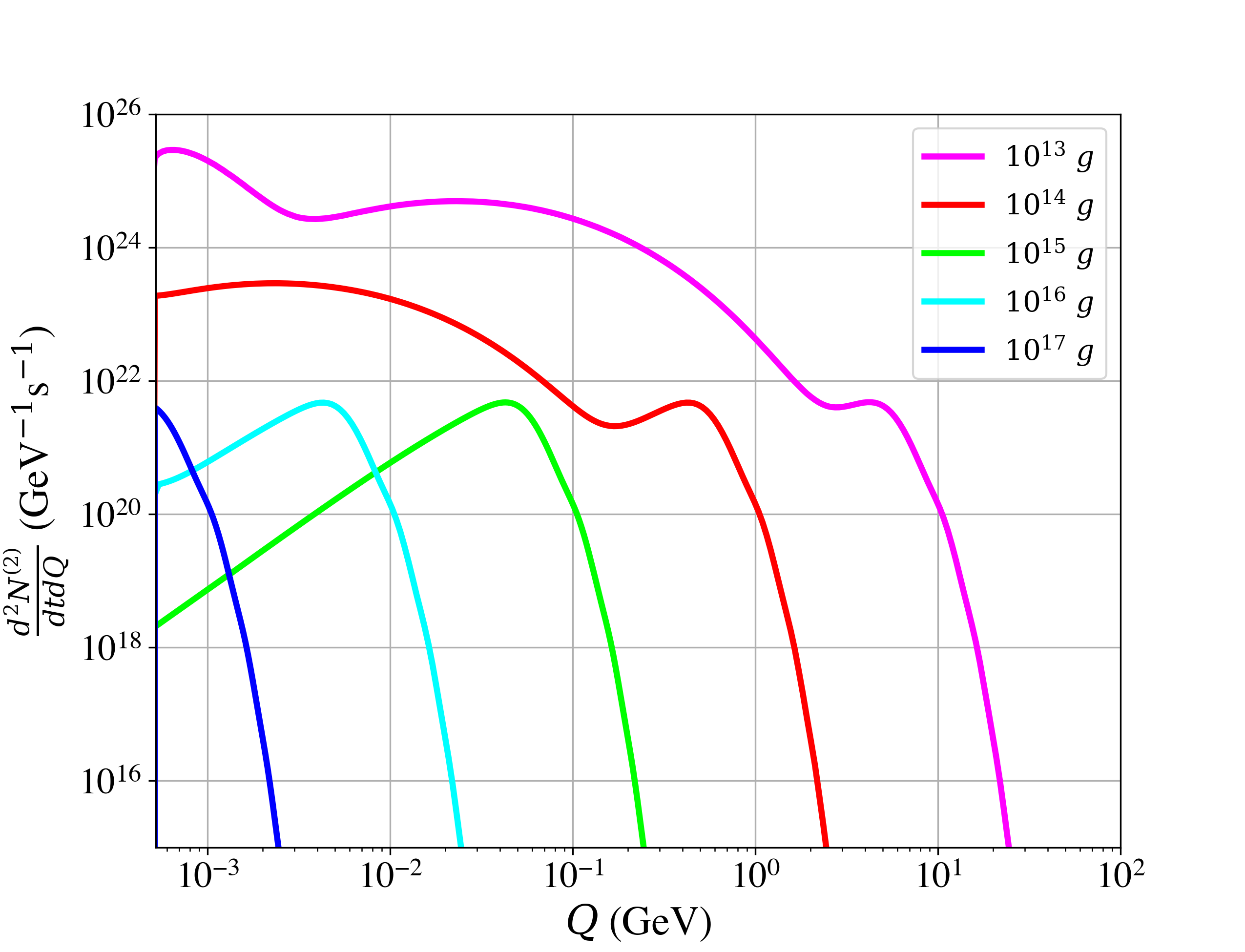}
\caption{\justifying Secondary positron emission spectra for PBHs in the positron-producing mass range. Note that the x-axis plots total positron energy $Q$ and thus cuts off at the rest mass $Q = 0.511$ MeV. PBHs with masses $M\gtrsim10^{14} \ {\rm g}$ have secondary spectra well-approximated by their primary spectra because they emit few short-lived heavier particles.}
\label{fig:PSComparison}
\end{figure}

\subsubsection{\label{sec:SecondarySpectra}Secondary Emission}
Due to their stability, electrons, positrons, photons, neutrinos, protons, and antiprotons provide the only measurable signatures of Hawking radiation from a PBH transiting past a detector with an impact parameter on the order of an astronomical unit (AU). We note that PBHs should, in principle, produce additional long-lived particles during the final explosive phase of their lives, including isotopes of hydrogen and helium and their anti-nuclei in equal numbers \cite{fisher_primal_2025}. 

The total emitted fluxes of stable particles, referred to as \textit{secondary emission spectra}, have contributions from their individual primary emission spectra (if the particles are fundamental), from decays of unstable primary particles, and from the hadronization and fragmentation of quark-gluon jets. This approach was pioneered by MacGibbon and Webber \cite{macgibbon_quark-_1990}, who argued that a black hole will emit primary particles which appear elementary at its temperature scale and that these primary particles will form composite particles and decay to stable, detectable secondary particles. 

The secondary emission spectrum for particle species $i$ is calculated via:
\begin{equation}
    \label{eqn:SecondaryEmisison}
    \frac{d^2N_i^{(2)}}{dt dQ} = \int_0^{\infty} \sum_j \frac{d^2N_j^{(1)}}{dt dQ'} \frac{dN_i^j}{dQ} dQ',
\end{equation}
where we sum over all primary Hawking particles $j$ and weight by the differential branching ratios $dN_i^j(Q', Q)/dQ$ \cite{arbey_blackhawk_2019}.

The numerical calculation of secondary emission spectra is carried out with \texttt{BlackHawk v2.2} \cite{arbey_blackhawk_2019, arbey_physics_2021}, which implements the particle physics codes \texttt{PYTHIA} \cite{sjostrand_introduction_2015} and \texttt{hazma} \cite{coogan_hazma_2020} to compute the branching ratios $N_i^j$ due to jets and decays. \texttt{PYTHIA} treats the emission of quarks and gluons as fundamental, and thus is only applicable for computing the secondary spectra of PBHs with masses $M \lesssim 10^{14}$ g, because such PBHs are hotter than the QCD scale (approximately 200 MeV). Within the mass range $M \gtrsim 10^{14} \text{ g }$, PBHs are too cold to emit quarks directly, since such quarks would be at energies below the pion rest mass, so for PBHs with masses in the range $10^{14} \, {\rm g} \lesssim M \lesssim 10^{17} \, {\rm g}$ we compute the secondary emission spectra with \texttt{hazma}, which treats pions as fundamental degrees of freedom and does not compute quark or gluon primary emission. For masses $M\gtrsim 10^{17} {\rm g}$, secondary positron spectra from \texttt{PYTHIA} and \texttt{hazma} converge, and we use \texttt{PYTHIA} as our afterburner. We note that integrated positron fluxes computed with \texttt{PYTHIA} and \texttt{hazma} differ by at most a few percent for PBHs with $M \gtrsim 10^{14} \, {\rm g}$. The primary and secondary \texttt{hazma} spectra for a PBH with $M = 10^{15} \, {\rm g}$ are shown in Fig.~\ref{fig:SpectraComparison}. 

The secondary emission spectra are considered the \textit{instantaneous} particle emission spectra from a PBH; but, in principle, these spectra are modulated by various energy-loss processes as the emitted particles propagate from the PBH to the detector. The positron flux as measured by a distant detector would be modified by processes including scattering off the interstellar (or interplanetary) medium, synchrotron radiation, photon pair-production, and bremsstrahlung. We ignore these ``tertiary'' contributions to the Hawking emission spectra due to the relatively short length scales, $\mathcal{O}(1\, {\rm AU})$, and low positron energy scales, $Q\lesssim 1 \, {\rm GeV}$, considered here.

\subsection{\label{sec:PBHMassFunctions}Distribution and Evolution of PBH Masses} 

Constraints on the PBH dark matter fraction $f_{\rm PBH}$ were originally derived assuming that all PBHs formed with some mass $\bar{M}$, thus obeying a ``monochromatic'' mass function, which is generally regarded as unphysical \cite{carr_primordial_2017,bellomo_primordial_2018,gow_accurate_2022}. Instead, in this section we introduce more realistic extended PBH mass functions, which incorporate physical features of the process by which PBHs form. Moreover, the PBH mass function will evolve over time, as individual PBHs undergo Hawking evaporation. Hence it is imperative to evolve such mass functions to the present day when evaluating constraints on $f_{\rm PBH}$. The time-evolution of the PBH mass function inevitably introduces a power-law tail for small masses, independent of whether the mass function at the time of PBH formation included such a feature. Incorporating such extended mass functions yields a modest tightening of constraints on the asteroid-mass window \cite{carr_primordial_2017,bellomo_primordial_2018,carr_constraints_2021,gorton_how_2024}.

\subsubsection{PBH Mass Functions}

At the time $t_i$ of PBH formation, the normalized PBH differential mass density function $\psi(M_i)$, referred to as the \textit{mass function}, is defined in relation to the differential number density distribution $dn_{\rm PBH}/dM$ via
\begin{equation}
    \label{eqn:PsiDef}
    \psi(M_i) = \frac{1}{\rho_{{\rm PBH},i}}\frac{d\rho_{ { \rm PBH}, i}}{dM_i} = \frac{M_i}{\rho_{ {\rm PBH},i}}\frac{dn_{ {\rm PBH}, i}}{dM_i} ,
\end{equation}
where $\psi(M_i)$ satisfies $\int dM_i \, \psi(M_i) = 1$ and determines the shape of the initial distribution. Here $n_{ {\rm PBH}, i}$ is the comoving PBH number density at formation time. 

PBHs form from the gravitational collapse of primordial curvature perturbations, in a process known as ``critical collapse'' \cite{choptuik_universality_1993,Evans:1994pj,Niemeyer:1999ak,Gundlach:2007gc,Musco:2008hv}. Upon re-entering the Hubble radius, perturbations with amplitude above a critical threshold will induce collapse, with the resulting PBH mass given by
\begin{equation}
    M_i = \kappa M_H (t_i) \vert \bar{\cal C} - {\cal C}_c \vert^\nu ,
    \label{MiCC}
\end{equation}
where $M_H (t_i)$ is the mass contained within a Hubble volume at time $t_i$, ${\cal C} = 2 G \delta M (r) / r$ is the compaction as a function of areal radius $r$ \cite{Shibata:1999zs,Harada:2023ffo}, $\bar{\cal C}$ is the spatial average of the compaction over a Hubble radius, ${\cal C}_c \simeq 0.4$ is the threshold for PBH formation \cite{Musco:2018rwt,Escriva:2019phb}, and $\kappa$ is an ${\cal O} (1)$ dimensionless constant whose value depends on the spatial profile of ${\cal C} (r)$ and on the averaging procedure \cite{Escriva:2021aeh}. The universal scaling exponent $\nu$ depends on the equation of state of the fluid that undergoes collapse; numerical studies find $\nu = 0.36$ for a radiation fluid \cite{Evans:1994pj,Niemeyer:1999ak,Musco:2008hv,Escriva:2021aeh}. 

Equation (\ref{MiCC}) implies that PBHs can form with arbitrarily small masses, if the (averaged) compaction is close to the critical threshold. Meanwhile, causality restricts PBHs to form with masses no larger than the Hubble mass at the time of formation. In fact, the typical mass at the peak of the distribution at the time of formation satisfies $\bar{M} (t_i) \simeq 0.2 \, M_H (t_i)$ \cite{carr_primordial_1975,niemeyer_near-critical_1998,Green:1999xm,Kuhnel:2015vtw}. These features of the critical collapse process generically yield PBH mass distributions that have a power-law tail for masses $M_i < \bar{M} (t_i)$ and an exponential cut-off for masses $M_i \gtrsim \bar{M} (t_i)$. 

Ref.~\cite{gow_accurate_2022} finds that these features of the PBH formation process are most accurately captured by the generalized critical collapse (GCC) parameterization of the PBH mass function. The normalized GCC mass function is defined as
\begin{equation}
    \label{eqn:GCCinit}
    \begin{split}
    \psi_{\rm GCC}&(M_i| \mu, \alpha, \beta) =  \\ &\frac{\beta}{\mu}\frac{1}{\Gamma((\alpha+1)/\beta)}\left( \frac{M_i}{\mu}\right)^{\alpha}\exp\left[ -\left(\frac{M_i}{\mu}\right)^{\beta}\right],
    \end{split}
\end{equation}
where $\mu$ is a location parameter with dimensions of mass, $\alpha > 1$ controls the power-law scaling of the low-mass tail, and $\beta > 0$ controls the high-mass exponential cutoff. The GCC mass function peaks at
\begin{equation}
    \label{eqn:MbarGCC}
    \bar{M}_{\rm GCC}(\mu, \alpha, \beta) = \mu \left( \frac{\alpha}{\beta}\right)^{1/\beta}.
\end{equation}
The original critical-collapse (CC) mass function corresponds to $\alpha = \beta = \nu^{-1}$.

\begin{figure}[]
    \centering
    \includegraphics[width=0.95\textwidth]{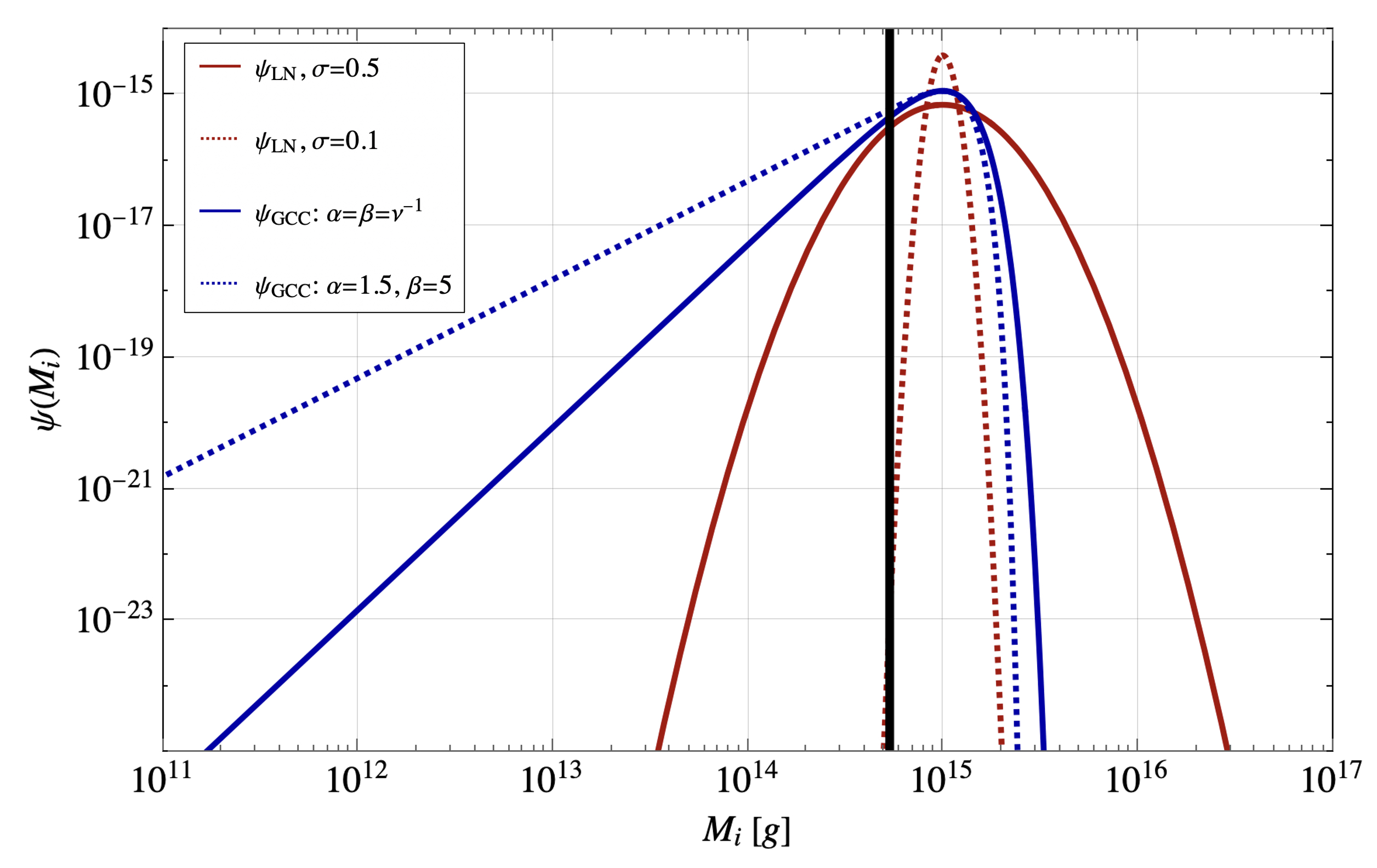}
    \caption{\justifying Comparison of GCC (blue) and LN (red)  formation-time mass functions with peaks located at $\bar{M} (t_i) =10^{15} \, {\rm g}$. The vertical black line indicates the universe-lifetime cutoff mass $M_*$ given in Eq.~(\ref{Mstar}). The main difference between the two parameterizations is the presence or absence of the small-mass tail.}
    \label{fig:FormationMFs}
\end{figure}

It has also been common in the literature to consider a simple log-normal (LN) distribution for the initial mass function \cite{carr_primordial_2020, kannike_single_2017,gorton_how_2024}:
\begin{equation}
    \label{eqn:LNinit}
    \psi_{\rm LN}(M_i|\mu, \sigma) = \frac{1}{\sqrt{2 \pi} \sigma M_i} \exp\left[ - \frac{\log{(M_i/\mu)}^2}{2 \sigma^2}\right].
\end{equation}
This model depends on two parameters: $\mu$, with dimensions of mass which sets the peak location of $M_i\psi(M_i)$, and $\sigma$, a dimensionless width parameter. The LN mass function $\psi_{\rm LN}(M_i)$ peaks at mass
\begin{equation}
    \label{eqn:MbarLN}
    \bar{M}_{\rm LN}(\mu, \sigma) = \mu e^{-\sigma^2}.
\end{equation}
Although the LN distribution depends on one fewer parameter than the GCC distribution, it omits the small-mass power-law tail, which comes directly from the critical-collapse relationship for $M_i$ of Eq.~(\ref{MiCC}). See Fig.~\ref{fig:FormationMFs}. As we will see, the small-mass tail of the initial mass function can play an important role when considering detection prospects for Hawking emission from PBHs near the lower edge of the asteroid-mass range.

\label{sec:PBHMassFunctions}
\subsubsection{PBH Mass Loss}
\label{sec:PBHMassLoss}
The mass of an individual PBH will evolve over time as it emits Hawking radiation \cite{macgibbon_quark-_1990, macgibbon_quark-_1991}:
\begin{equation}
\begin{split}
    \label{eqn:MassEvolution}
    \frac{dM}{dt} &= -\sum_j g_j\int_{m_{\rm eff}, j}^{\infty}dQ \, \frac{Q}{c^2}\frac{d^2N^{(1)}_j}{dtdQ}\\
    &=-A \, \frac{f(M)}{M^2} , 
    \end{split}
\end{equation}
where the sum is taken over all fundamental Standard Model particles, $d^2N^{(1)}_j/dtdQ$ is the primary Hawking spectrum for particle $j$, and $A=5.19\times10^{25} \,\, {\rm g}^3\,{\rm s}^{-1}$. 

\begin{figure}[]
    \centering
    \includegraphics[width=0.95\textwidth]{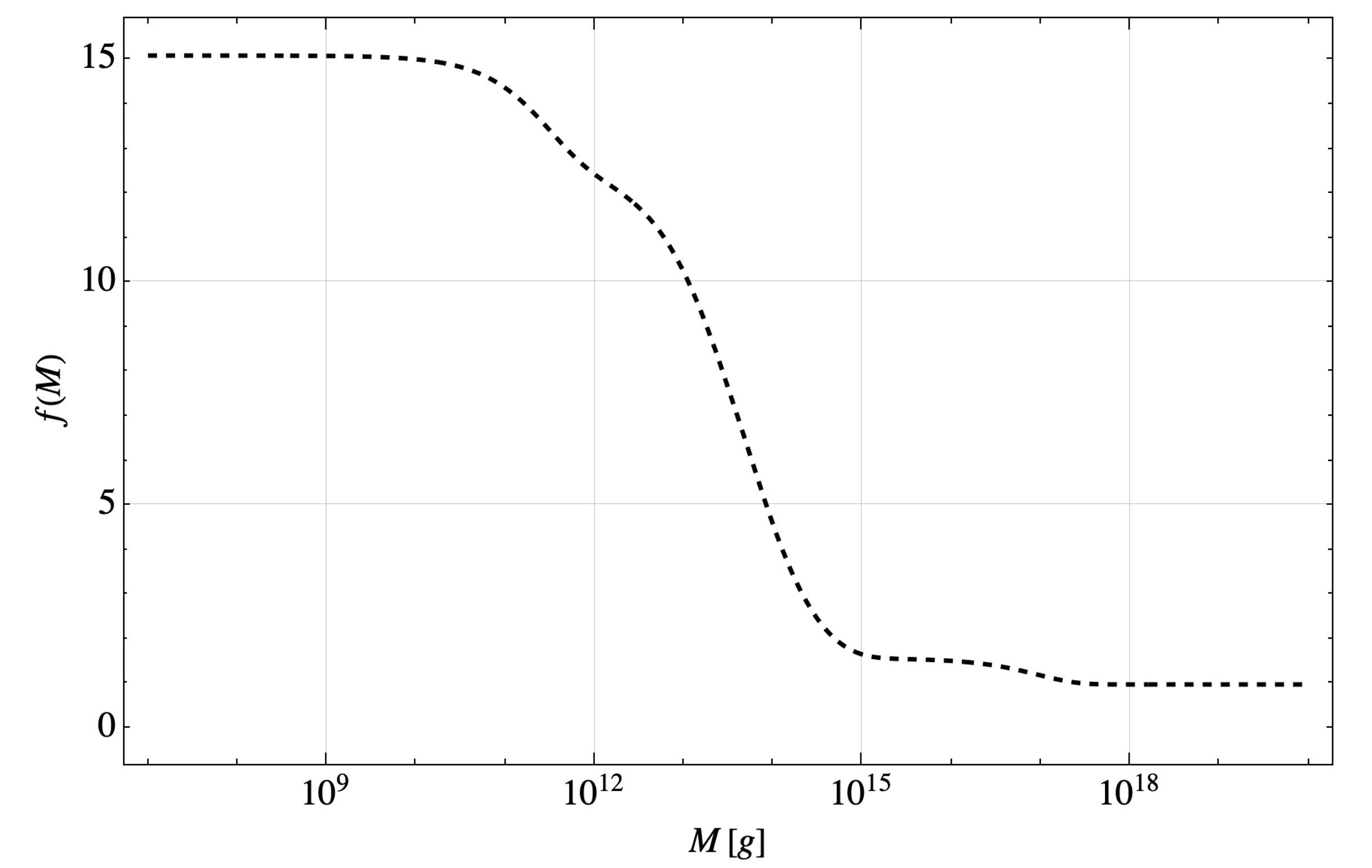}
    \caption{\justifying The Page factor $f (M)$ as a function of PBH mass as defined in Eq.~(\ref{eqn:PageFactor}). We have normalized the Page factor such that $f(M)=1$ for PBHs emitting only photons and neutrinos, which are assumed massless. PBH mass $M$ here indicates mass at time of particle emission.}
    \label{fig:PageFactor}
\end{figure}

The function $f(M)$, referred to as the \textit{Page factor}, quantifies the total number of quantum degrees of freedom that can be emitted by a PBH of mass $M$. The value of $A$ is chosen to normalize the Page factor such that $f(M)=1$ for PBHs emitting only photons and neutrinos (which are treated as massless). We follow the method of Ref.~\cite{macgibbon_quark-_1991} and define the Page factor as
\begin{equation}
    \label{eqn:PageFactor}
    f(M) \simeq \mathcal{P}_{\gamma}+\mathcal{P}_{\nu}+\sum_j \mathcal{P}_j\exp\left(-\frac{M}{M_j} \right),
\end{equation}
where the Page coefficients $\mathcal{P}_j$ and the quantum degrees of freedom (DOF) $g_j$ can be found in Table 2 of the Supplemental Materials of Ref.~\cite{klipfel_ultra-high-energy_2025}. The characteristic mass $M_j$ is defined as the PBH mass at which the emitted power for species $j$ peaks at the effective particle mass $m_{{\rm eff}, j}$. Note that, unlike Ref.~\cite{macgibbon_quark-_1991}, we include terms for the $W$, $Z$, and Higgs bosons and use updated particle masses. The inclusion of the exponential factors in $f(M)$ is a phenomenological way to smoothly ramp up the total number of DOF as particle production ``turns on'' for each new species when $T\approx m_{{\rm eff}, j}$. The Page factor is plotted in Fig.~\ref{fig:PageFactor}.

We can write down an approximate analytical solution to Eq.~(\ref{eqn:MassEvolution}) by solving the equation
\begin{equation}
    \int \frac{M^2}{f(M)} dM = -\int A \, dt + C ,
\end{equation}
where $C$ is an integration constant. We integrate by parts on the right-hand side:
\begin{equation}
\label{eqn:intparts}
    \begin{split}
    \int \frac{M^2}{f(M)} dM & = \frac{1}{3}\frac{M^3}{f(M)}-\int \frac{1}{3}M^3 \frac{d}{dM}f(M)^{-1} dM\\ 
    & = \frac{1}{3}\frac{M^3}{f(M)}+\int \frac{1}{3}\frac{M^3}{f(M)^2} \frac{df(M)}{dM}dM \\
    & \approx \frac{1}{3}\frac{M^3}{f(M)},
    \end{split}
\end{equation}
where we have used the assumption that $f(M)$ varies slowly---i.e., that $df(M)/dM \approx 0 \, \forall \, M$. This assumption holds within the regime of interest to us because the second term on the right-hand side of the second line of Eq.~(\ref{eqn:intparts}) is at least an order of magnitude smaller than the first for all masses $M\geq10^{12} \, {\rm g}$.

Upon using the initial condition $M(t_i=0) = M_i$, we then recover a relation between $M$ and $M_i$ at time $t$:
\begin{equation}
\label{eqn:MMiRelation}
    \frac{1}{f(M)}M^3 = -3At + \frac{1}{f(M_i)}M_i^3.
\end{equation}
(See also Ref.~\cite{Cai:2021fgm}.) It is common to make the further assumption that $f(M) = f(M_i)$, which allows one to solve directly for $M(t)$  \cite{mosbech_effects_2022}:
\begin{equation}
    \label{eqn:MApprox}
    M(t, M_i) \approx \left(-3Af(M_i)t + M_i^3 \right)^{1/3},
\end{equation}
for its inverse:
\begin{equation}
    \label{eqn:MiApprox}
    M_i(t, M) \approx \left(3Af(M)t + M^3 \right)^{1/3},
\end{equation}
and for the approximate PBH lifetime \cite{carr_cosmological_1976, macgibbon_quark-_1991}:
\begin{equation}
    \label{eqn:LifetimeApprox}
    \tau(M_i) \approx \frac{1}{3A} \frac{M_i^3}{f(M_i)}. 
\end{equation}
However, for $M_i \sim M_*$ we see that $f(M_0) \approx 10 f(M_i)$, which implies that this assumption breaks down and Eqs.~(\ref{eqn:MApprox})--(\ref{eqn:LifetimeApprox}) become inaccurate for initial masses near $M_*$. (Note that $M_0$ indicates the present-day mass.) A more accurate result for $M(t, M_i)$ can be obtained by numerically solving Eq.~(\ref{eqn:MMiRelation}) using $f(M)$ as defined in Eq.~(\ref{eqn:PageFactor}). Solving numerically for the value of $M_i$ such that $M(t_0, M_i) = 0$, where the present age of the universe is $t_0 \equiv 13.787 \pm 0.020 \, {\rm Gyr}$ \cite{planck_collaboration_planck_2020}, gives the universe-lifetime cutoff mass: 
\begin{equation}
M_* = (5.364 \pm 0.002)\times10^{14} \, {\rm g}.
\label{Mstar}
\end{equation}
PBHs that form with $M_i = M_*$ would just be completing their Hawking evaporation process today.

\subsubsection{Present-Day Mass Functions}
\label{sec:PresentDayPsi}

The simplest way to construct an approximate present-day PBH mass function is to truncate the formation-time mass function $\psi (M_i)$ for $M_i \leq M_*$. Assuming that all PBHs with $M_i < M_*$ have evaporated, one could then define a present-day mass function $\psi(M,t_0)$ via $\psi (M, t_0) \propto \psi (M_i) \, \Theta (M_i - M_*),$ where $\Theta (x)$ is the Heaviside step function. Such a distribution could be normalized by requiring $\int_{M_*}^\infty dM \, \psi (M, t_0) = 1$. In practice, for PBH mass functions that peak at $\bar{M} \gg M_*$, one may neglect the evolution of the PBH population over time and simply set $\psi (M_i) \simeq \psi (M)$, which is commonly done in the literature.

However, for mass functions that are peaked near the lower-mass end of the asteroid-mass range and that have extended small-mass tails, this truncation procedure is insufficient to accurately describe the present-day PBH distribution. Such a procedure would neglect a real population of PBHs with masses below the cutoff $M_*$ today, which exist due to mass-loss by heavier PBHs over cosmological timescales. To accurately model present-day mass functions that peak near the asteroid-mass range, we must therefore use the mass-loss formalism discussed in Section \ref{sec:PBHMassLoss} to evolve the formation-time mass functions to the present day \cite{Martin:2019nuw,mosbech_effects_2022,gorton_how_2024,cang_21-cm_2022,Boluna:2023jlo,klipfel_ultra-high-energy_2025}.

We first introduce the normalized \textit{PBH number distribution function}  \cite{mosbech_effects_2022, cang_21-cm_2022}:
\begin{equation}
    \label{eqn:PhiDef}
    \phi(M_i) \equiv \frac{1}{n_{{\rm PBH}, i}}\frac{dn_{{\rm PBH},i}}{dM_i}.
\end{equation}
This function is defined such that $\phi(M_i)dM_i$ is the fraction of PBHs with mass $M_i$ at formation time, and satisfies $\int dM_i\,\phi(M_i) = 1$.

We can derive an exact relation between $\phi$ and $\psi$ by combining Eqs.~(\ref{eqn:PhiDef}) and (\ref{eqn:PsiDef}):
\begin{equation}
    \label{eqn:PhiPsiRelation}
    \phi(M_i) = {\cal C} \frac{1}{M_i}\psi(M_i),
\end{equation}
where
\begin{equation}
    {\cal C} \equiv \left[\int_0^{\infty} \frac{dM_i}{M_i}\psi(M_i) 
    \right]^{-1}.
\end{equation}

\begin{figure}[t]
    \centering
    \includegraphics[width=0.95\textwidth]{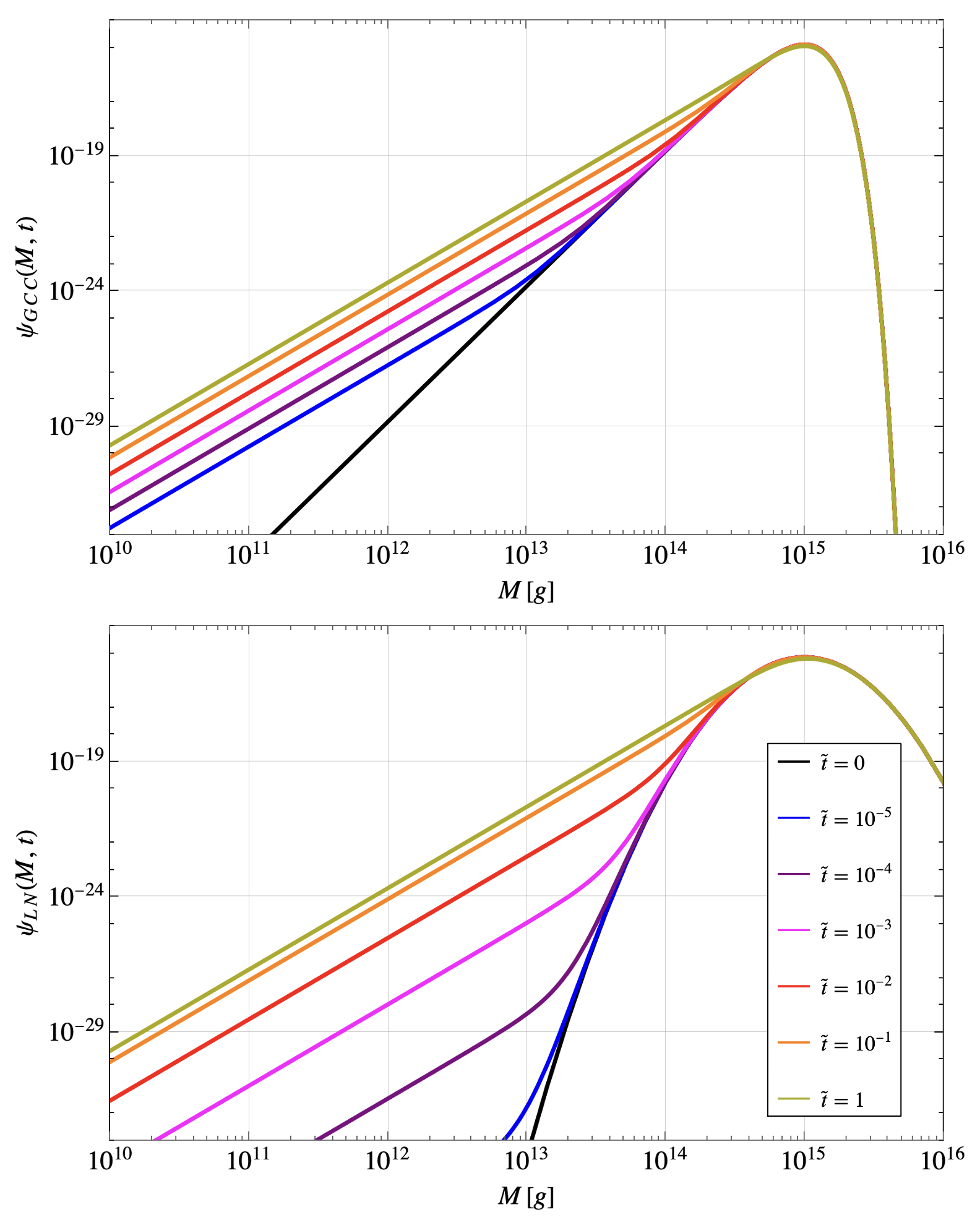}
    \caption{\justifying Time-evolution of the mass function $\psi_{\rm GCC} (M, t)$ with $\alpha = 5$ and $\beta = 2$ (top) and $\psi_{\rm LN} (M, t)$ with $\sigma=0.5$ (bottom).
    Both functions are peaked at $\bar{M}_i=10^{15} \, {\rm g}$. The curves are plotted for different times $\tilde{t}$ in units of the current age of the universe, so that $\tilde{t}=1$ corresponds to the present day.}
    \label{fig:AnApprox}
\end{figure}

\noindent The peak mass of the distribution, $\bar{M}$, is a good approximation for the normalization constant ${\cal C}$ for sharply-peaked distributions. Therefore, following Ref.~\cite{klipfel_ultra-high-energy_2025} we can relate
\begin{equation}
    \phi(M_i) \approx  \frac{\bar{M}}{M_i}\psi(M_i),
\end{equation}
where $\bar{M}$ is given by Eq.~(\ref{eqn:MbarGCC}) or (\ref{eqn:MbarLN}), depending on the form of the initial-time PBH mass function. 

To derive a general expression for $\psi(M,t)$ at some time $t>t_i$, we take advantage of the conservation of the total number of PBHs within a certain mass bin in the present day. The number $N$ of PBHs with mass in some range $[M_1, \, M_2], \,\, M_1>0,$ at time $t$ is given by the integral over the function $\phi(M)$:
\begin{equation}
    \label{eqn:PhiCoordTransInt}
    \begin{split}
    N & = n_{{\rm PBH}, i}\int_{M_1}^{M_2}dM \, \phi(M, t) \\
    & = n_{{\rm PBH}, i} \int_{M_i(M_1, t)}^{M_i(M_2, t)}dM \, \phi(M(M_i, t), t_i)\left[ \frac{dM}{dM_i} \right]^{-1},
    \end{split}
\end{equation}
where, in the second line, we implemented a time-dependent change of coordinates $M\to M_i$ defined by $M = M(M_i,t)$ with Jacobian $|dM/dM_i|$, which can be derived from Eq.~(\ref{eqn:MMiRelation}). Equating the integrands allows us to define the time-dependent PBH number distribution function as
\begin{equation}
    \label{eqn:GenPhiMt}
    \phi(M, t) = \phi(M_i(M, t), t_i) \frac{dM_i(M,t)}{dM},
\end{equation}
where $M_i(M,t)$ is the inverse of our coordinate transformation $M(M_i,t)$. Note that we have introduced new notation to explicitly encode the time-dependence: $\phi(M)$ at time $t $ is $ \phi(M,t)$ and $\phi(M_i) \equiv \phi(M_i, t_i)$. 

Ref.~\cite{mosbech_effects_2022} uses Eq.~(\ref{eqn:GenPhiMt}) and the relation $M_i(M,t)$ defined in Eq.~(\ref{eqn:MiApprox}) as a coordinate transformation to derive an approximate analytical expression for $\phi(M,t)$:
\begin{equation}
    \label{eqn:PhiTimeApprox}
    \begin{split}
    \phi(M,t)  & \approx M^2\phi\left(M_i(M,t), t_i\right) \\
    & \,\,\,\,\,\,\times \left(M^3+3Atf(M_i(M,t))\right)^{-2/3}.\\
    \end{split}
\end{equation}

\noindent We can then use Eqs.~(\ref{eqn:PhiPsiRelation}) and (\ref{eqn:GenPhiMt}) to construct a general expression for the time-dependent mass function $\psi(M,t)$ \cite{cang_21-cm_2022}:
\begin{equation}
    \label{eqn:GenPsiMt}
    \psi(M,t) = \frac{M}{M_i(M,t)}\psi(M_i(M, t), t_i) \frac{dM_i(M,t)}{dM}.
\end{equation}
Substituting Eq.~(\ref{eqn:MiApprox}) as before to construct an approximate form of the time-dependent mass function yields
\begin{equation}
    \label{eqn:PsiTimeApprox}
    \begin{split}
    \psi(M,t) & \approx \frac{M^3}{M_i(M,t)} \psi\left(M_i(M,t), t_i\right) \\
    & \,\,\,\,\, \times \left(M^3+3Atf(M_i(M,t))\right)^{-2/3}. \\
    \end{split}
\end{equation}
We will refer to Eqs.~(\ref{eqn:PhiTimeApprox}) and (\ref{eqn:PsiTimeApprox}) as the \textit{analytical approximations} for $\phi$ and $\psi$. Figure~\ref{fig:AnApprox} plots the analytical approximations of $\psi_{\rm GCC}(M,t)$ and $\psi_{\rm LN}(M,t)$  at various times $\tilde{t} \equiv t/t_0$ for $\bar{M}=10^{15} \, {\rm g}$, with GCC parameters $\alpha=5$ and $\beta=2$ and LN parameter $\sigma=0.5$, where $t_0$ is the present age of the universe.

\begin{figure}[t]
    \centering
    \includegraphics[width=0.95\textwidth]{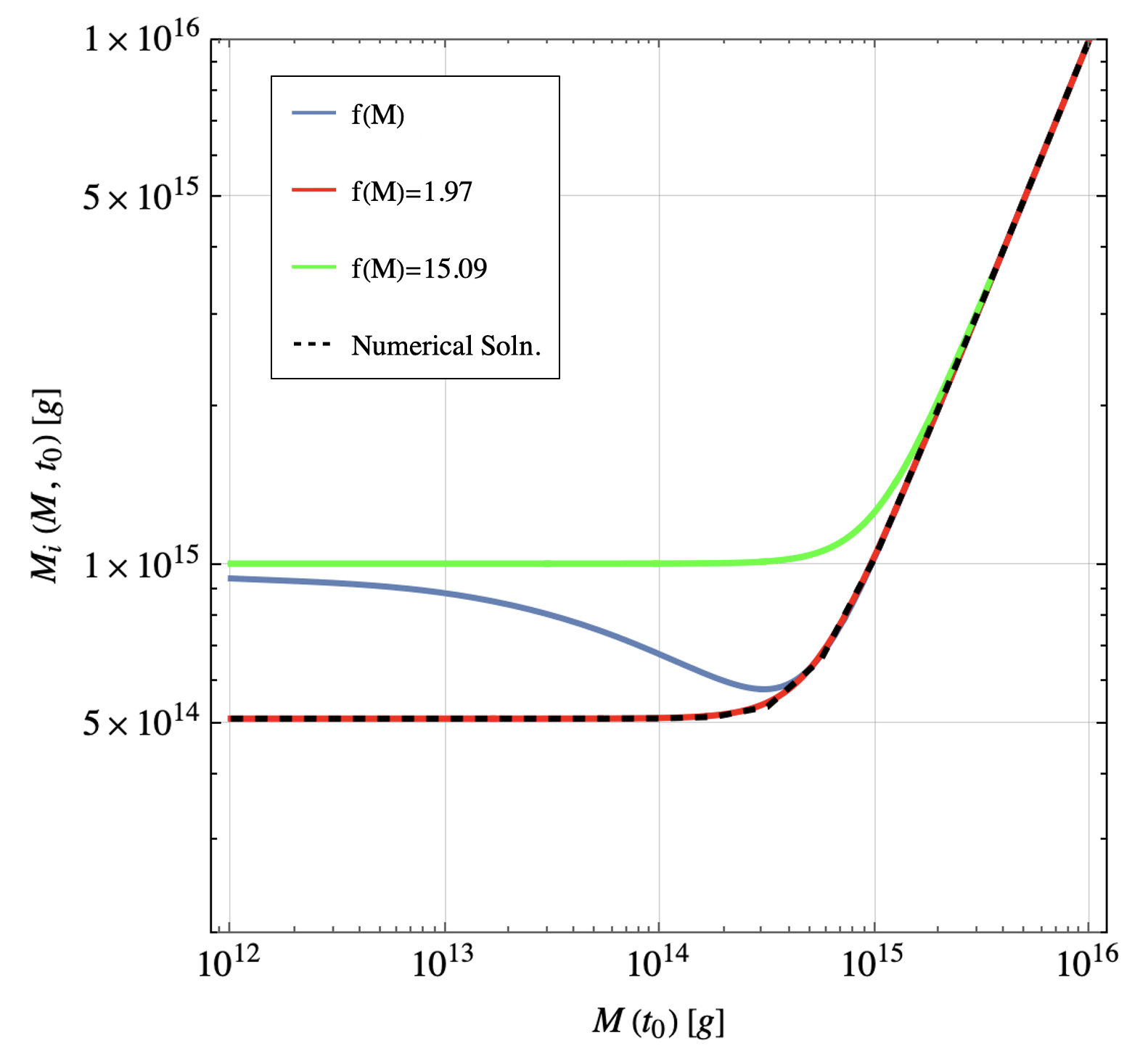}
    \caption{\justifying Approximate and exact numerical relations between formation-time mass $M_i$ and present-day mass $M (t_0)$. The solid curves plot the approximate expression $M_i(M,t)$ from Eq.~(\ref{eqn:MiApprox}) with different assumptions for the Page Factor: explicit $M$-dependence according to Eq.~(\ref{eqn:PageFactor}) (blue), $f(M) = f(M_*) =1.97$ (red), and $f(M) = \lim_{M\to0}f(M)=15.09$ (green). The dashed black curve is a numerical solution to Eq.~(\ref{eqn:MMiRelation})  given the exact relation between $M_i$ and $M$, which is very well approximated by setting $f(M)=f(M_*)$. }
    \label{fig:MiMRelations}
\end{figure}

It's relevant to note that the approximate relations between $M$ and $M_i$ given by Eqs.~(\ref{eqn:MApprox}) and (\ref{eqn:MiApprox}), though convenient, result in an unphysical behavior in the mass evolution. Figure~\ref{fig:MiMRelations} plots $M_i(M,t)$ from Eq.~(\ref{eqn:MiApprox}) (blue), which does not display the monotonic behavior expected due to PBH evolution being ``well-ordered'' \cite{cang_21-cm_2022}. If two PBHs have masses $M_1$ and $M_2$ at time $t_i$ satisfying $M_1<M_2$, then we expect their masses to satisfy $M_1 (t) < M_2 (t)$ at any future time $t>t_i$---that is to say, that mass-ordering is preserved. Figure~\ref{fig:MiMRelations} shows the result from solving Eq.~(\ref{eqn:MMiRelation}) numerically (black) and also plots Eq.~(\ref{eqn:MiApprox}) assuming that the Page factor is independent of mass: $f(M) = f(M_*) =1.97$ (red) and $f(M) = \lim_{M\to0}f(M)=15.09$ (green). We find that (at least for masses $M (t_0) \geq 10^{12} \, {\rm g}$) it is a much better approximation of the true behavior to use $M_i(M,t)$ from Eq.~(\ref{eqn:MiApprox}) with $f(M)\approx 1.97$ than to include the mass-dependence of the Page factor. 

All present-day mass functions used in this article are computed using Eq.~(\ref{eqn:PsiTimeApprox}) with $\psi(M_i,t_i)$ given by either Eq.~(\ref{eqn:GCCinit}) (GCC) or Eq.~(\ref{eqn:LNinit}) (LN), and with $M_i(M,t)$ given by Eq.~(\ref{eqn:MiApprox}) with $f(M)=1.97$ fixed. 

As shown in Fig.~\ref{fig:AnApprox}, the present-time mass functions $\psi (M,t)$ feature power-law small-mass tails for both the GCC and LN distributions. In the GCC case, depending on the value of $\alpha$, the small-mass tail of the time-evolved mass function can feature a more gentle slope than the formation-time mass function. More striking, the LN distribution develops a comparable small-mass tail to the GCC distribution over time, even though the formation-time LN distribution includes no such tail at all. When estimating detection rates for phenomena like cosmic rays emitted via Hawking emission from small-mass PBHs, it is therefore critical to time-evolve the PBH mass functions to the present day. 

\subsection{\label{sec:LocalDMDistribution}Local Dark Matter Distribution}

The rate at which PBHs transit through the inner Solar System depends on the present-day PBH mass function, the local DM density, and the PBH velocity distribution relative to the Sun. In this section we discuss our assumptions for local PBH density and velocity distributions. 

\subsubsection{\label{DMDensityProfile}Dark Matter Density Profile}

The visible Milky Way lies inside a large, approximately spherical dark matter halo \cite{davis_evolution_1985, franx_ordered_1991}. The DM density distribution of this halo $\rho_{\rm DM}(\vec{r})$ is established through galactic rotation curves, and is typically modeled as a Navarro-Frenk-White (NFW) profile \cite{navarro_structure_1996}, which is characterized by two parameters: a reference density $\rho_0$ and scale radius $r_0$. 

We use the modified NFW-like density profile from Refs. \cite{binney_modelling_2017, posti_mass_2019}:
\begin{equation}
    \label{eqn:NFW}
    \rho_{\rm DM}(r, z) = \frac{\rho_0}{L(1+L)^2}\exp{\left[-\left(\frac{L r_0}{r_{vir}}\right)^2 \right]},
\end{equation}
where $r$ and $z$ are cylindrical coordinates in the Galactocentric frame and the best-fit parameters from Ref. \cite{posti_mass_2019} are:
$$L=\sqrt{\left(\frac{r}{r_0}\right)^2 + \left(\frac{z}{qr_0}\right)^2},$$
\begin{equation}
\begin{split}
    &\rho_0=0.0196 \, M_{\odot}\text{pc}^{-3},\\
    &r_0=15.5 \, \text{kpc},\\
    &r_{vir}=287,\\
    &q=1.22.\\
\end{split}
\end{equation}
Other halo models in recent literature include a generalized Navarro-Frenk-White (gNFW) profile or the Einasto profile \cite{salas_estimation_2019, ou_dark_2023}.

The local dark matter density in the neighborhood of our Solar System, $\rho_{\rm DM}^{\odot}$, can be estimated by Jeans analysis of stellar motions \cite{kafle_shoulders_2014} or by constructing galactic density models and rotation curves \cite{ou_dark_2023, weber_determination_2010, salas_estimation_2019}. (See Ref.~\cite{read_local_2014} for a review of local DM density measurements and techniques.) Measured values for $\rho_{\rm DM}^{\odot}$ generally fall between $0.2-1.5 \ \text{GeV/cm}^3$ \cite{read_local_2014}. 
Evaluating Eq.~(\ref{eqn:NFW}) at the location of the Solar System, $r_{\odot}=8.3 \, {\rm kpc}$ and $z_\odot \simeq 0$, gives the local DM density: 
\begin{equation}
\label{eqn:SolarDMdensity}
\rho_{\rm DM}^{\odot} = 0.0155 \, M_{\odot} \, {\rm pc}^{-3} = 0.589 \, \text{GeV/cm}^3. 
\end{equation}
We will use this value throughout our analysis. 

The local PBH differential number density will therefore be given by
\begin{equation}
    \label{eqn:LocaldndM}
    \frac{dn}{dM}=f_{\rm PBH}\,\rho_{\rm DM}^{\odot}\frac{1}{M}\psi(M,t_0),
\end{equation}
where $\psi(M,t_0)$ is the 
present-day mass function computed according to Section \ref{sec:PresentDayPsi}.

\subsubsection{\label{DMVelocityProfile}Dark Matter Velocity Profile}

We treat our local region of the DM halo as a volume of isotropically distributed, non-interacting PBHs with a fixed number density given by the integral of Eq.~(\ref{eqn:LocaldndM}). We neglect any potential effects due to PBH interactions, clustering, or binaries \cite{carr_observational_2024}. 

\begin{figure}[t]
    \centering
    \includegraphics[width=0.95\textwidth]{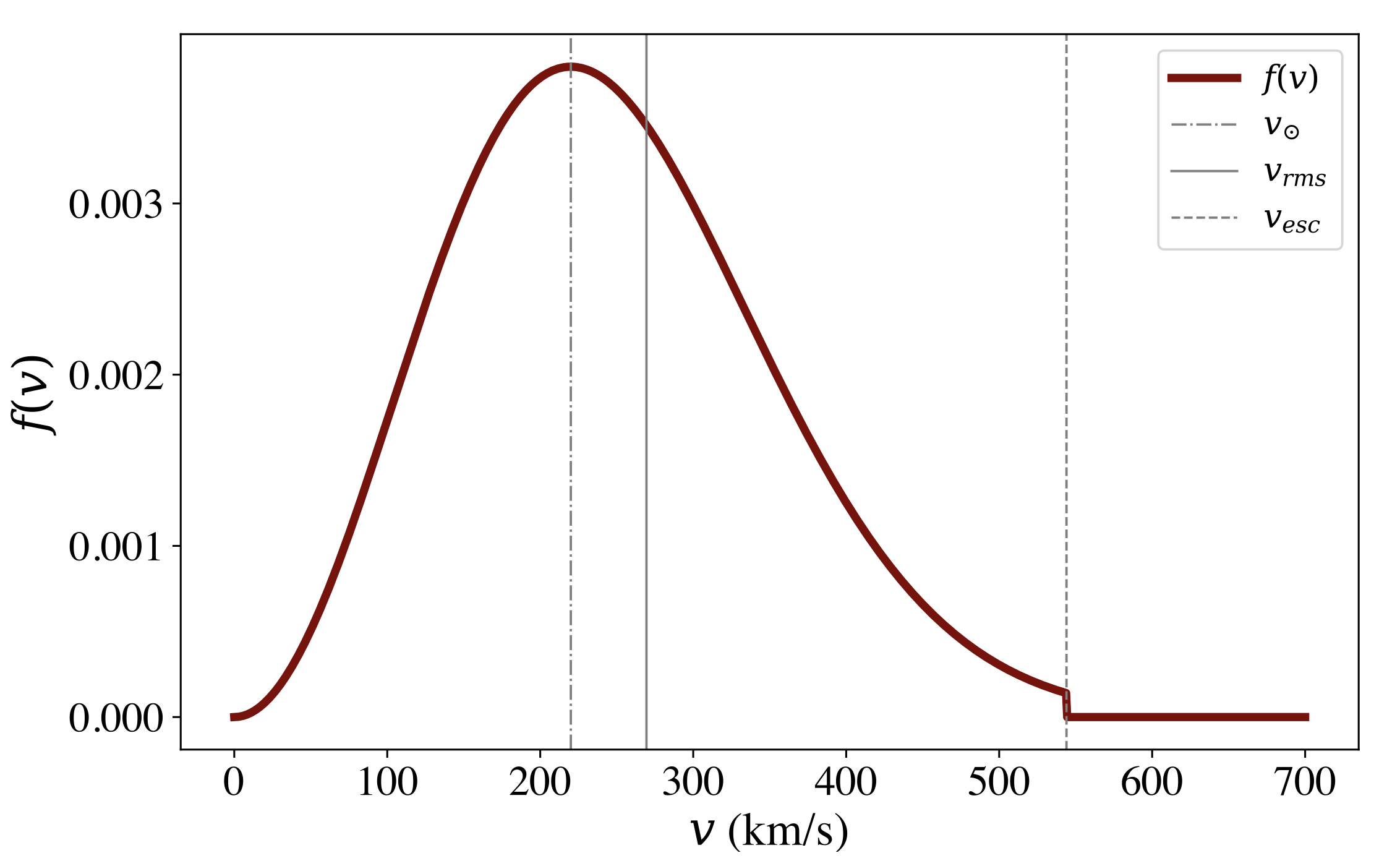}
    \caption{\justifying Maxwellian velocity profile for dark matter located $8$ kpc from the center of the Milky Way. $v_{\odot}$ is azimuthal velocity of the Sun in the galactic plane and $v_{\text{rms}}$ is the root mean square velocity for the Maxwellian distribution. The distribution drops to zero at the galactic escape velocity $v_{\rm esc} \approx 544 {\rm km/s}$. }
    \label{fig:VelProfile}
\end{figure}

This population of PBHs will obey a Maxwellian velocity distribution \cite{cerdeno_particle_2010, choi_impact_2014}:
\begin{equation}
    \label{eqn:Maxwellian}
    f(v) = \frac{4 f_0}{\sqrt{\pi}} \left( \frac{3}{2}\right)^{3/2} \frac{v^2}{v_{\text{rms}}^3} \exp{\left[ - \frac{3}{2}\frac{v^2}{v_{\text{rms}}^2} \right]}; \ \ v < v_{\text{esc}}.
\end{equation}
We assume that the Sun is located at $r_{\odot} = 8.0\pm0.5$ kpc, which corresponds to $v_{\odot}=220\pm20$ km/s \cite{cerdeno_particle_2010}. The velocity dispersion is related to the Sun's azimuthal velocity in the galactic plane via $v_{\text{rms}} = \sqrt{3/2}\, v_{\odot} \approx 270$ km/s \cite{choi_impact_2014}. Truncating Eq.~(\ref{eqn:Maxwellian}) at the galactic escape velocity $v_{\text{esc}}=544$ km/s sets the normalization constant $f_0=1.00684$. This DM velocity profile is plotted in Figure \ref{fig:VelProfile}.

More complicated models of local DM velocity distributions include modified Maxwellians \cite{oh_velocity_1997}, Eddington inversion methods \cite{christy_dark_2024}, and results from numerical simulations. (See Ref.~\cite{choi_impact_2014} for a review.) Furthermore, observational measurements hint at possible sources of local DM density and velocity anisotropies due to astrophysical phenomena like the Nyx stellar stream \cite{wang_high-resolution_2023, donlon_ii_local_2022}.  This object is a high-eccentricity, prograde stellar stream within the Milky Way thin disk which may be the remnant of an ancient collision between a dwarf galaxy and the Milky Way \cite{wang_high-resolution_2023}. If Nyx is the result of a dwarf galaxy collision, it would contribute to the local dark matter substructure and affect direct detection experiments \cite{bruch_detecting_2009}. The impact on PBH transit rates caused by such modifications to the local DM velocity profile is beyond the scope of this paper. 

In the following analysis we assume PBHs obey Eq.~(\ref{eqn:Maxwellian}) and that the out-of-the-galactic-plane ($v_z$) and radial ($v_r$) velocity components are negligible in comparison to the azimuthal velocity $v_{\phi}$. Because the relative velocity between the Earth and the Sun is $\sim 0.1 v_{\text{rms}}$, we also assume that the contribution from Earth's orbit around the Sun to $f(v)$ is negligible with respect to $v_{\text{rms}}$. This amounts to a purely azimuthal PBH ``wind'' with $v_{\phi} \approx \bar{v} = 246$ km/s flowing past the Earth, analogous to the ``WIMP wind'' discussed in association with WIMP direct detection experiments. 

\subsection{\label{ExptlConsiderations}Detector Geometry}

In this section we discuss the technical considerations relevant to measuring Hawking radiation from low-mass PBHs transiting past Earth with a detector stationed in low-Earth orbit (LEO). Figure \ref{fig:PSComparison} shows secondary emission spectra for select PBHs in the positron-producing mass range, $M_* \leq M \leq 5 \times10^{17}$g. Because the particles emitted during these close encounters would only propagate over length scales of ${\cal O} (1 \, {\rm AU})$ rather than over kiloparsecs, we can neglect modifications due to energy loss, scattering, and diffusion effects typically relevant for computing cosmic ray spectra. However, the shape and energy range of the measured signal will be modified by the geomagnetic field and the specific detector geometry.

The Alpha Magnetic Spectrometer (AMS) is a candidate experiment to detect Hawking radiation from transiting PBHs, with over 10 years of high-precision time-series positron flux data \cite{aguilar_temporal_2023}. Other detectors with potentially useful datasets include FERMI-Lat, which could also study coincident time-dependent $\gamma$-ray and $e^{\pm}$ signatures, PAMELA, and Voyager-1 (which would not be subject to the geomagnetic field constraints discussed in following subsections).

We focus on positron signatures in this article because PBHs emit $e^+$ and $e^-$ at equal rates but positrons have a significantly lower astrophysical background. Section \ref{sec:PositronBkgd} address the complexities of the positron background and the effects of solar modulation on our signature. Time-series analysis of $e^{+}$ fluxes is uniquely possible with AMS data because of the high charge separation resolution made possible by the AMS 0.15 T permanent magnet. The electron charge confusion fraction is below one in $10^4$ for energies below $30 \, {\rm GeV}$ \cite{aguilar_alpha_2021}. Other charged lepton Hawking radiation constraints, such as those placed by Voyager-1 data \cite{boudaud_voyager_2019}, are computed with $(e^- + e^+)$ fluxes. 

\subsubsection{\label{sec:GeomagField}Geomagnetic Field}

Measuring charged cosmic rays with a detector in LEO requires an understanding of the magnetosphere, the volume around Earth dominated by its magnetic field. Earth's magnetic field at radius $r$, latitude $\theta$, and longitude $\lambda$ can be described by the spherical harmonic expansion of a scalar potential $\Psi$:
\begin{equation}
    \label{eqn:GeomagneticPotential}
    \begin{split}
    \Psi(r, \theta, \lambda) = & a \sum_{l=1}^{\infty}\sum_{m=0}^{l}  \left(\frac{a}{r} \right)^{l+1} \Biggl[ C_l^m \cos{(m \lambda)} \\ &
     + S_l^m \sin{(m \lambda)} \Biggr] P_l^m(\cos{\theta}),
     \end{split}
\end{equation}
which satisfies $\vec{B} = - \nabla \Psi$ and $\nabla^2 \Psi = 0$ \cite{lowes_magnetic_2011, alken_international_2021}. In Eq.~(\ref{eqn:GeomagneticPotential}), $a$ is a reference radius and $C_l^m$ and $S_l^m$ are often referred to as \textit{Gauss coefficients} which can be fit to geomagnetic field data using an appropriate reference radius \cite{huder_cov-obsx2_2020}. 

The geomagnetic field can be approximated by the $l=1$ dipole term of Eq.~(\ref{eqn:GeomagneticPotential}) with magnetic moment $7.7\times 10^{22}$Am$^2$ \cite{alken_international_2021}. The magnetic moment is not static spatially or temporally and has fluctuated and reversed over the course of Earth's history. Recent measurements show the dipole moment has been decreasing since at least the year 1600 \cite{korte_centennial_2012}. Archeomagnetic and paleomagnetic data from core samples, thermoremnant magnetization, and archaeological samples show complex variation of the dipole moment over large time-scales \cite{korte_centennial_2012, panovska_one_2019}, and evidence that the geomagnetic field reverses over uncorrelated intervals of tens of thousands to millions of years in a statistically random process \cite{gaffin_analysis_1989, phillips_spectral_1976}. The dipole axis is not parallel to the Earth's rotation axis, and the magnetic north dipole is currently located near $86.5^o$ latitude \cite{alken_international_2021}. The locations of the magnetic poles drift with time, and are projected by the most recent International Geomagnetic Reference Field report to travel at an average rate of $40$km/yr from 2020-2025 \cite{alken_international_2021}. Since 1900, the magnetic North pole has traveled northward by approximately $15^o$ latitude \cite{alken_international_2021}.

Assuming a dipole field geometry, the Stoermer approximation can be used to estimate the \textit{geomagnetic cutoff} rigidity $R_c$ \cite{stormer_polar_1955}:
\begin{equation}
    R_c = \frac{\mathcal{M} \cos^4{\theta}}{r^2 \left[1 + \sqrt{1 - \sin{\epsilon} \sin{\xi} \cos^3{\theta} } \right]^2}.
\end{equation}
This cutoff is the minimum \textit{rigidity} (momentum per unit charge) necessary for a particle to reach a point in the magnetosphere from infinity, and it is generally a function of particle path, location, and geomagnetic field \cite{kress_modeling_2015}. $R_c$, with units of MV, is a function of radius $r$ in cm, magnetic latitude $\theta$, azimuthal angle $\xi$, zenith angle $\epsilon$, and magnetic moment $\mathcal{M}$ in $\text{G}\cdot \text{cm}^3$, all defined in Ref.~\cite{smart_review_2005}. The geomagnetic cutoff rigidity ranges from about $15$ GV at the equator to $0$ GV at the magnetic poles, where the tangential component of the field vanishes. Sub-GeV charged cosmic rays can therefore only be detected by an experiment in LEO when its orbit takes it near the magnetic poles---where charged particles can penetrate to lower altitudes. This reduces both the measured positron signal and background and significantly reduces the proton background at low energies for an experiment in LEO. We include the effect of the geomagnetic cutoff in our simulations of time-dependent positron signals from PBH transits measured by AMS in section \ref{sec:PBHsignature}.

\subsubsection{\label{sec:AMSDetector}AMS Detector}

AMS is a precision particle physics experiment aboard the International Space Station (ISS) equipped with a $0.15$ T permanent magnet and six detector subsystems designed to measure the charge $q$, momentum $p$, velocity $\beta\equiv v/c$, and rigidity $R\equiv p/q$ of fundamental particles and nuclei \cite{aguilar_alpha_2021}. The detector stack includes 9 silicon tracker layers, a Transition Radiation Detector (TRD), upper- and lower- Time of Flight (TOF) counters, an Anti-Coincidence Counter (ACC) surrounding the magnet bore, a Ring Imaging Cherenkov (RICH) detector, and an Electromagnetic Calorimeter (ECAL). AMS has been operating since 2011 and to date has published data from $3.4\times10^6$ positrons with energies between $500$ MeV and $1$ TeV \cite{aguilar_temporal_2023, aguilar_towards_2019}.  

AMS is subject to the ISS orbital trajectory in LEO with an inclination of $51.6^o$ and a period of $90$ minutes. As discussed in the previous subsection, because the geomagnetic north dipole has moved to such a high latitude over the past decades, experiments like AMS---even with its high orbital inclination---are quite constrained by the geomagnetic cutoff. AMS is only able to measure low rigidity (sub-GV) particles at the most extreme northern and southern latitudes of its orbit. As a result, observation time for 1 GV positrons is only on the order of 1000 seconds per day \cite{aguilar_temporal_2023}.

Other constraints on the energy range and precision of low-energy positron measurements with AMS come from the AMS Level-1 trigger, particle track curvature, detector resolution, and background rejection efficiency. The Level-1 trigger for electrons and positrons requires a coincidence of hits on all 4 TOF layers and a minimum ECAL energy deposition \cite{aguilar_alpha_2021}. Considering that the upper and lower TOF counters (with two layers each) are separated by 1.25 m \cite{Bindi:2010zzb}, and accounting for the thickness of the RICH ($0.47 \, {\rm m}$) \cite{Giovacchini:2014ora} and the ECAL ($0.167 \, {\rm m}$) \cite{aguilar_alpha_2021}, the total distance from the top TOF layer to the bottom of the ECAL is $\sim 2 \, {\rm m}$. 
Therefore, an estimate for the minimum radius of curvature such that an incident positron can hit both TOF counters and the ECAL is about $1 \, {\rm m}$, which translates to a minimum detectable positron energy of $\sim$50 MeV. 

There is a reported drop in the AMS Level-1 trigger efficiency starting at around 3 GeV and reaching an efficiency of $\sim 80\%$ for 500 MeV electrons \cite{aguilar_alpha_2021}. Furthermore, ECAL energy resolution rises above the $10\%$ level for $e^{\pm}$ below $1$ GeV and proton rejection rates from the TRD and ECAL drop by several orders of magnitude for sub-GeV positrons \cite{aguilar_alpha_2021}. The combination of these detector properties and the geomagnetic cutoff has led the AMS collaboration to only publish on positrons with energy $Q \geq Q_{\rm min}=500$ MeV, for which statistics are relatively high.  

In Sec.~\ref{sec:Results} we study the feasibility of using published time-series $e^+$ flux data from AMS to detect time-dependent positron signatures from nearby PBH transits. We further discuss applying this analysis technique to potential future positron datasets. The lower the positron energy that the data includes, the larger the space of PBH mass function models that we can probe. 

\section{\label{sec:Results}Results}
\subsection{\label{sec:TransitRates}PBH Transit Rates}

\begin{figure}[t]
    \centering
    \includegraphics[width=0.8\textwidth]{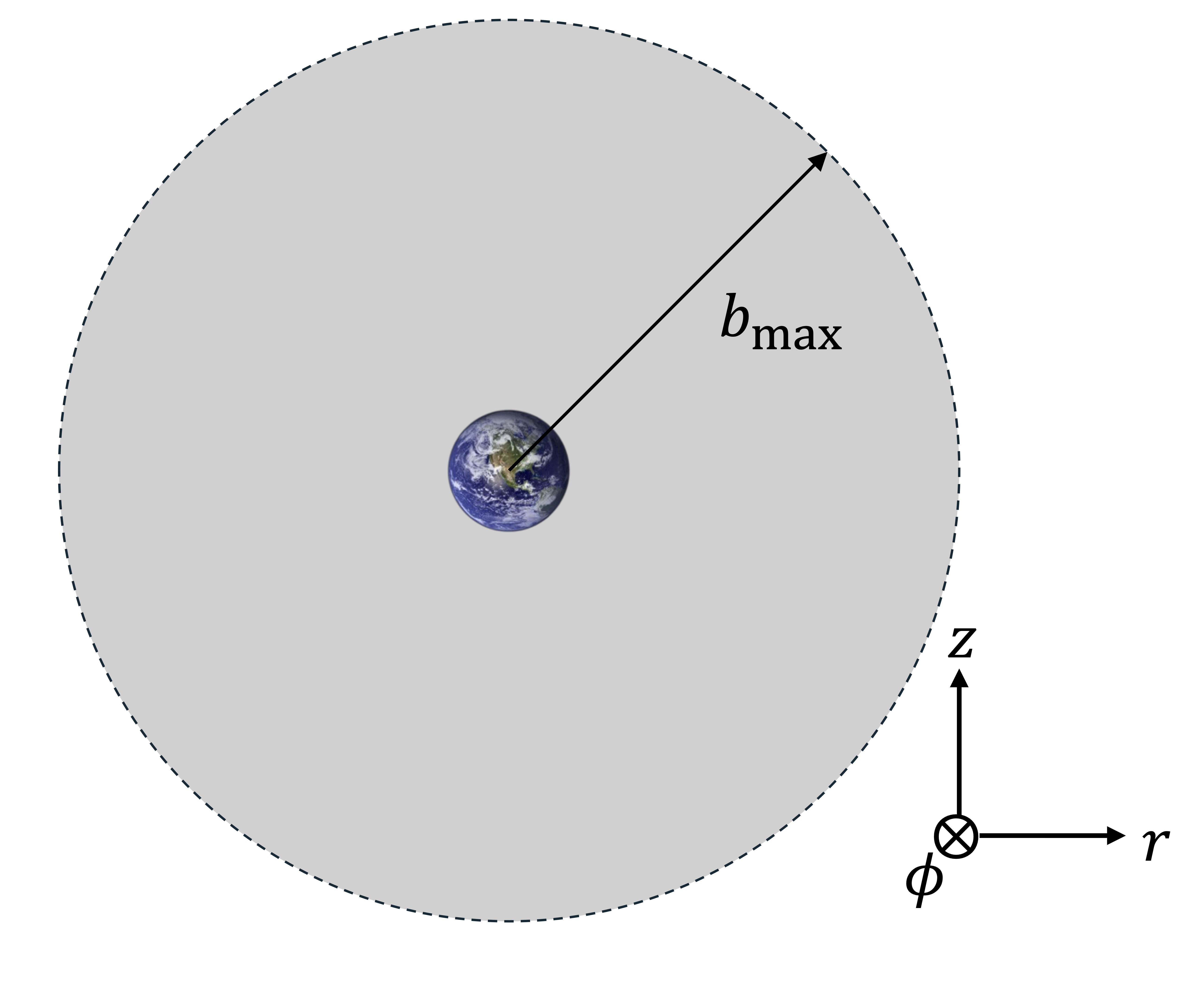}
    \caption{\justifying Schematic of the transit geometry discussed in Section \ref{sec:TransitRates}. PBHs traveling in the $\hat{\phi}$ direction, parallel to the galactic plane, will transit past Earth within some impact parameter $b_{\text{max}}$ at a rate given by Eq.~(\ref{eqn:ImpactParamIntegral}).}
    \label{fig:TransitSchematic}
\end{figure}

We estimate the number of PBHs that will transit past the Earth per year with impact parameter $b$. We let $f_{\rm PBH}=1$ because we aim to generate new constraints independent of existing evaporation constraints.  A schematic of the transit geometry is depicted in Figure \ref{fig:TransitSchematic}.  We assume that PBHs with masses distributed according to a present-day mass function $\psi(M,t_0)$ are traveling parallel to the galactic plane with azimuthal velocity $v_{\phi}$ distributed according to Eq.~(\ref{eqn:Maxwellian}) and that $v_{\phi} \gg v_r, v_z$. 

The differential transit rate for PBHs of mass $M$, velocity $v$, and impact parameter $b$ is:
\begin{equation}
    \label{eqn:ImpactParamDifferential}
    \frac{d^3\Phi_{\rm transit}}{dMdvdb} = \frac{dn}{dM} (v f(v)) (2 \pi b).
\end{equation}
Here we have assumed that the mass and velocity distributions are locally isotropic. We may then estimate the number of PBHs with masses up to some maximum mass $M_{\rm max} (t_0)$ today that should transit past the Earth per unit time with impact parameter $b \leq b_{\rm max}$. We average over the velocity distribution $f (v)$, integrate over the mass range $0 \leq M (t_0) \leq M_{\rm max} (t_0)$, and integrate up to $b_{\rm max}$, which might be mass-dependent:
\begin{equation}
    \label{eqn:ImpactParamIntegral}
    \begin{split}
    \bar{\Phi}_{\rm transit}(M_{\rm max}, b_{\text{max}}) & = \rho_{\rm DM}^{\odot}\int_{0}^{M_{\rm max}} dM \,\frac{1}{M} \psi (M, t_0) \\ & \ \ \ \ \times \int_0^{v_{\rm esc}} dv \,v f(v) \int_0^{b_{\rm max}(M)}db \, 2\pi b , 
    \end{split}
\end{equation}
where the over-bar indicates averaging over the PBH velocity distribution and $\rho_{\rm DM}^{\odot}$ is the local dark matter mass density given by Eq. ~(\ref{eqn:SolarDMdensity}). Here we have used the definition of $dn/dM$ from Eq.~(\ref{eqn:LocaldndM}) with $f_{\rm PBH}=1$.

\begin{figure*}[t]
    \centering
    \includegraphics[width=.95\textwidth]{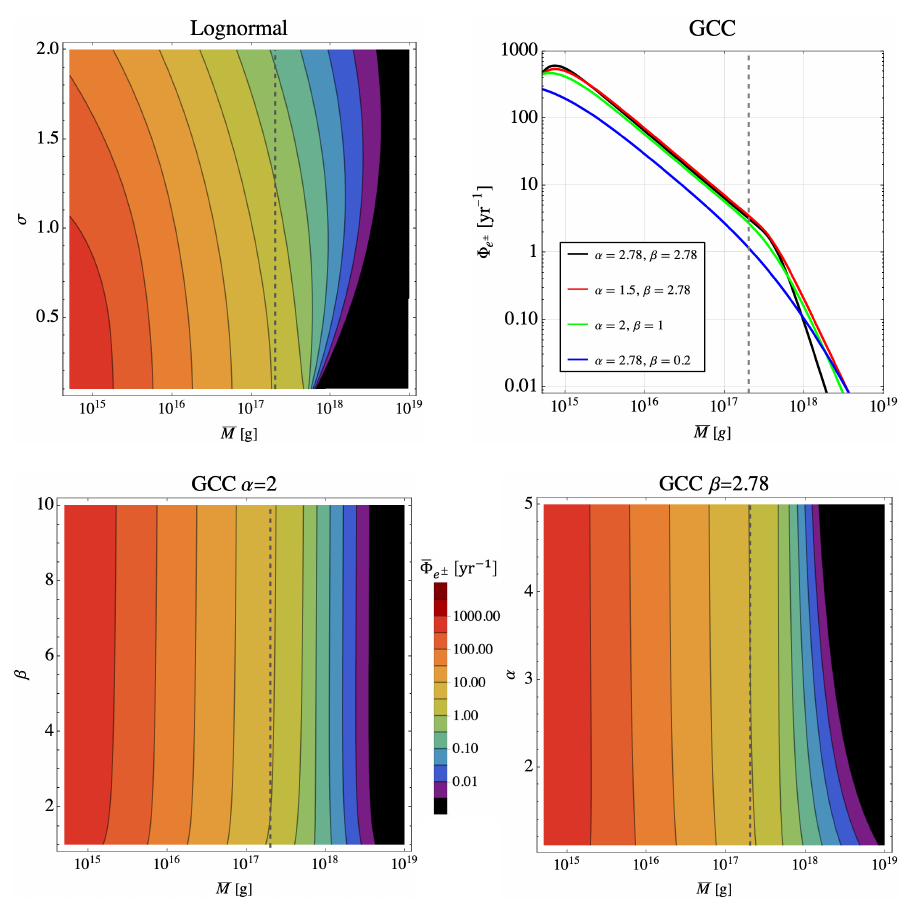}
    \caption{\justifying Number of expected positron-producing PBHs that pass within $1 \, {\rm AU}$ of Earth per year. Here we plot the yearly transit rate $\bar{\Phi}_{e^\pm}$ according to Eq.~(\ref{eqn:Phiepmdef}) with $b_{\rm max}=1 \, {\rm AU}$, $t=t_0$, and $f_{\rm PBH}=1$, as a function of the mass $\bar{M}_i$ at the peak of the formation-time PBH mass function, given in Eqs.~(\ref{eqn:MbarGCC}) and (\ref{eqn:MbarLN}) for the GCC and LN functions, respectively.  
    Rates for the LN mass function range from $0.005-1065$ yr$^{-1}$ for $0.1 \leq \sigma \leq 2.0$. The GCC PBH transit rate (top right) is plotted as a function of $\bar{M}_i$ and shown for three sets of parameters $\{\alpha, \beta\}$. Contour plots of the GCC PBH transit rate with $\alpha$ constant (bottom, left) and $\beta$ constant (bottom, right) indicate that the transit rate is only weakly sensitive to $\alpha$ and $\beta$ for $\bar{M}_i>M_*$. For both models, if $\bar{M}_i$ falls in the positron-producing mass range, $M_*\leq \bar{M}_i \leq 5\times10^{17}$g, the expected transit rates exceed $1 \, {\rm yr}^{-1}$, which implies that a decade of precision time-series data could probe large regions of parameter space. Vertical dashed lines at $\bar{M}_i=2\times10^{17} \, {\rm g}$ indicate the mass where current evaporation constraints on extended mass functions stop and $f_{\rm PBH} = 1$ is viable \cite{luque_refining_2024,gorton_how_2024}. Note that detectable positron signals could arise even for $\bar{M}_i > 5 \times 10^{17} \, {\rm g}$ because we have incorporated realistic, extended mass functions and evolved them from the time of PBH formation to today; hence a subpopulation of PBHs with $M (t_0) \leq 5 \times 10^{17} \, {\rm g}$, capable of emitting positrons today, would emerge from the small-mass tail of the underlying PBH mass function.    }
    \label{fig:TransitRate}
\end{figure*}

As noted above, only PBHs with present-day mass $M (t_0) \leq 5 \times 10^{17} \, {\rm g}$ will emit electrons and positrons. Hence to study the number of PBH transits that could yield detectable positrons today, we consider the quantity
\begin{equation}
    \bar{\Phi}_{e^\pm} (b_{\rm max}) \equiv \bar{\Phi}_{\rm transit} (5 \times 10^{17} \, {\rm g} , b_{\rm max}) .
    \label{eqn:Phiepmdef}
\end{equation}
Given how modest the mass-loss due to Hawking evaporation is over the current age of the universe for $M_i = 5 \times 10^{17} \, {\rm g} \gg M_* = 5.364 \times 10^{14} \, {\rm g}$, we incur an error of only $( M(t_0, M_i) - M_i ) / M_i = - 3.6 \times 10^{-10}$ for $M_i = 5 \times 10^{17} \, {\rm g}$, per Eq.~(\ref{eqn:MApprox}). Hence we may use the positron-emitting mass $M(t_0) \simeq M_i = 5 \times 10^{17} \, {\rm g}$ when evaluating $\bar{\Phi}_{e^\pm} (b_{\rm max})$.

The PBH transit rate $\bar{\Phi}_{e^\pm}(b_{\text{max}})$ is model-dependent and therefore a function of either the GCC model parameters $\mu$, $\alpha$, and $\beta$ or the LN model parameters $\mu, \sigma$. Figure \ref{fig:TransitRate} shows the yearly transit rate $\bar{\Phi}_{e^\pm}$ for $b_{\text{max}} = 1 \, {\rm AU}$ and $t=t_0$ as a function of the respective model parameters for both GCC and LN mass functions. 
For both models, we expect at least one transit per decade for $\bar{M}_i\lesssim 10^{18} \,{\rm g}$. It is encouraging that a cosmic ray experiment with a decade of data could potentially probe such a large space of models via a careful time-series analysis.

Given that both the GCC and LN parameter spaces admit such high transit rates for $f_{\rm PBH}=1$, one may wonder whether PBHs colliding with the Earth, Sun, or other Solar System bodies would provide better constraints on the dark matter fraction of low-mass PBHs than Hawking radiation signatures. Stellar capture of PBHs was first posited by Hawking \cite{hawking_gravitationally_1971}. Recent work estimates stellar capture rates \cite{Khriplovich:2009jz,Capela:2012jz,Capela:2013yf,Genolini:2020ejw,Lehmann:2020yxb,Lehmann:2022vdt,Caiozzo:2024flz,Tinyakov:2024mcy} and analyzes the evolution of ``Hawking stars'' with captured black holes at their centers \cite{oncins_primordial_2022, bellinger_solar_2023, caplan_is_2024}. Potential signatures of Hawking stars would be unique solar seismological signals or disappearing stars \cite{bellinger_solar_2023,Tinyakov:2024mcy}, or weak gravitational-wave signals from bound-PBH orbits within its host star \cite{DeLorenci:2025wbn}. Stellar capture rates are predicted to be quite low, however, and none of these signatures have yet been observed. Refs.~\cite{yalinewich_crater_2021, santarelli_possible_2025} discuss the morphology of craters from PBH impacts on rocky bodies and suggest that a study of lunar craters may provide PBH dark matter constraints. We estimate that, for a LN mass function with $\mu \in [M_*, 5\times10^{17} \text { g}]$ and $\sigma \in [0.1, 2]$ (the parameter space shown in Fig.~\ref{fig:TransitRate}), between about 1 and 4000 PBHs would have impacted the Earth over its 4.54 Gyr existence. Given that impact rates are so low and because the Earth is tectonically active, this is not a likely detection strategy. 

Searching for evidence of PBH transits through the significantly larger volume of the inner Solar System, however, allows high enough transit rates that we may see PBH candidates within the approximately decade-long lifetimes of existing cosmic ray experiments---even for models peaked in the unconstrained asteroid-mass range where number densities are lower.  Note that other recent analyses that study gravitational perturbations from PBH flybys also find comparable transit rates to those calculated in this study, but do not use time-evolved PBH mass functions \cite{tran_close_2023,Cuadrat-Grzybowski:2024uph}.

\subsection{\label{sec:PBHsignature}Time-Dependent PBH Signature}

The goal of this section is to develop an analysis technique to extract a PBH transit signal from time-series cosmic ray flux data, and to estimate an upper bound on detection efficiency for a given detector geometry when transit parameters $M$, $v$ and $b$ are known.  
We specifically perform numerical studies to determine how reliably the AMS experiment could detect time-dependent positron excesses corresponding to PBH transits with mass $M$ and impact parameter $b$. 

We first develop a Monte Carlo simulation to model the measured Hawking emission signal from a transiting PBH within a noisy positron background. Section \ref{sec:PBHSignals} derives analytically the estimated time-dependent positron count-rate measured by the AMS detector in orbit around the Earth during a PBH transit. In Sec.~\ref{sec:PositronBkgd} we construct our positron background using mean fluxes and temporal fluctuation amplitudes sampled from AMS data \cite{aguilar_towards_2019, aguilar_temporal_2023}. Then in Sec.~\ref{sec:MatchedFilter} we discuss low-SNR peak detection 
with matched filters. In Sec.~\ref{sec:Detectability} we develop a measure of ``detectability'' for a transit with parameters $M$ and $b$ and compute the \textit{maximum detectable impact parameter}, $b_{\text{max}}(M)$ for a PBH of mass $M$. 
Section \ref{sec:ModelsProbed} utilizes our simulation results to update the plots in Fig.~\ref{fig:TransitRate}  
by computing the number of \textit{detectable} positron-producing PBH transits per year as a function of model parameters for simulated positron datasets from each of three detector configurations.

The three detector configurations, with parameter values listed in Table \ref{table:DetConfigs}, are: (1) AMS aboard the ISS in low-Earth orbit, (2) AMS in geosynchronous orbit, and (3) a hypothetical detector located at a Lagrange point outside the geomagnetic field. 
For configuration (2), we posit that a future positron dataset extending down to approximately $Q_{\rm min}=50 \, {\rm MeV}$ could be collected by an instrument similar to AMS (or AMS itself) in geosynchronous orbit rather than from LEO. With a height $h\approx 36,000 \, {\rm km}$, an instrument in geosynchronous orbit would see a maximum Stoermer cutoff rigidity at equatorial latitudes of $R_c\approx48 \, {\rm  MeV}$ \cite{noauthor_geomagsphere_nodate, bobik_magnetospheric_2006, boschini_geomagnetic_2013, grandi_trajectory_2016}. Furthermore, as discussed in Sec.~\ref{sec:AMSDetector}, we estimate that based on AMS detector geometry, the minimum positron energy that would correspond to a detectable radius of curvature and an ECAL hit would be $\mathcal{O}(50 \,{\rm MeV})$. Similarly, for configuration (3) we consider a hypothetical detector of similar resolution and acceptance to AMS, but located even farther outside the Earth's magnetosphere---such as at a Lagrange point. Such a detector, with instrument modifications like a weaker magnetic field strength, could be sensitive to positrons with energies $Q\geq5 \, {\rm MeV}$ and could probe a much wider space of realistic PBH mass functions. We 
identify regions of model parameter space that could be explored with each dataset.

\begin{figure}[t]
\centering
\includegraphics[width=0.95\textwidth]{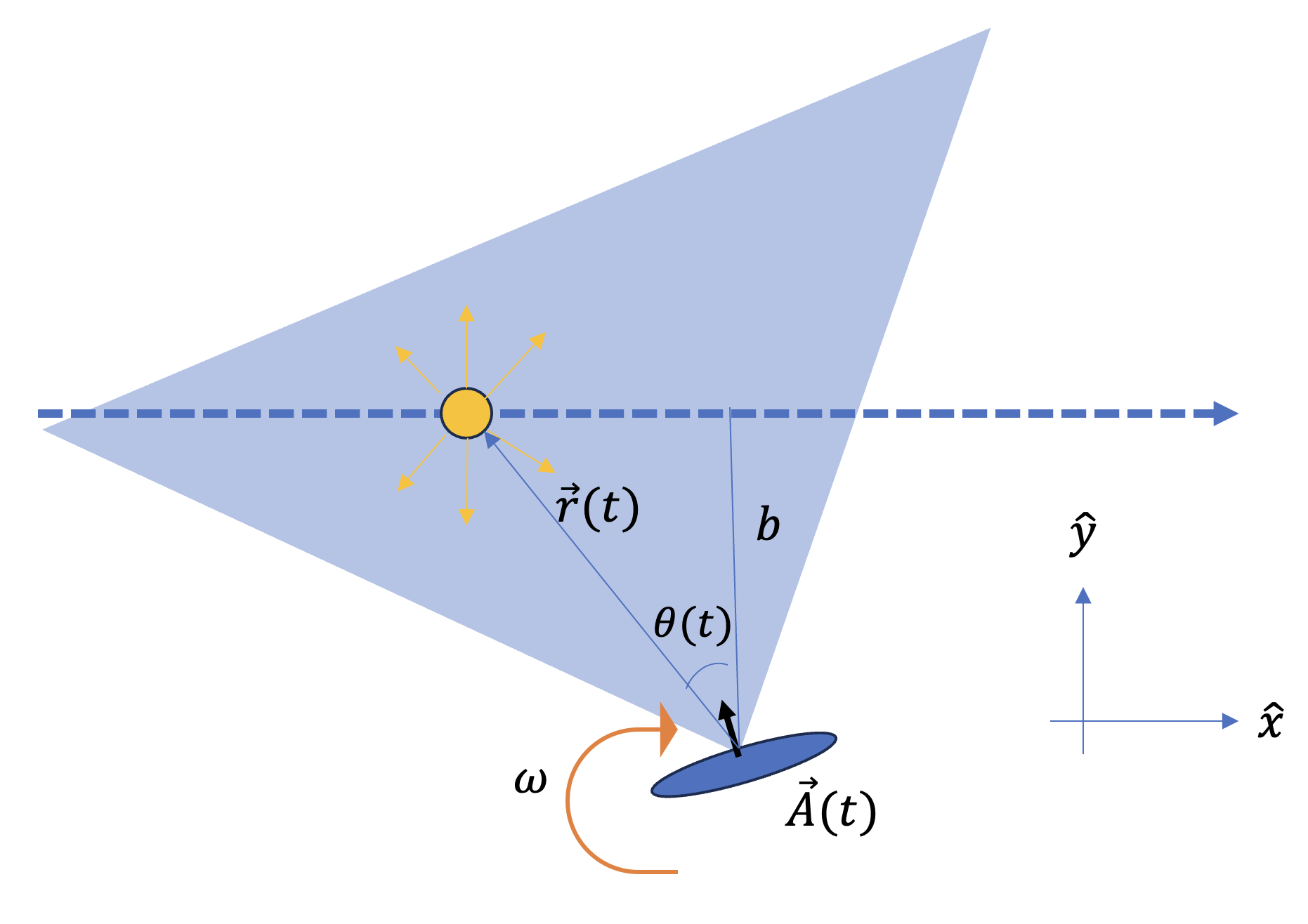}
\caption{\justifying Geometry for simulating the time-dependent $e^+$ signal from a PBH transit measured by a detector in orbit about the Earth. The blue field of view cone rotates with the detector area $\vec{A}(t)$ about a fixed point at frequency $\omega=2\pi/\tau$. We approximate the orbit as rotation about a fixed point because $b \gg R_{\mathTerra}+h$. The PBH (yellow circle) transits past on a linear trajectory with velocity $\bar{v} =246 \, {\rm km/s}$ while emitting positrons at a rate $\dot{N}(M)$ given by Eq.~(\ref{eqn:ToyModelPBHFlux}).}
\label{fig:TransitGeometry}
\end{figure}

\begin{table}[h!]
\caption{\label{table:DetConfigs}\justifying The three detector configurations simulated in this study, corresponding to AMS in low-Earth orbit, AMS in geosynchronous orbit, and a hypothetical detector located at a Lagrange point well outside the geomagnetic field. The parameter $\tau_{\rm obs}(Q)$ encodes the effects of the geomagnetic field via imposing latitude-dependent cutoff rigidities. For AMS in LEO, $\tau_{\rm obs}(Q)$ is interpolated form reported values and spans the range $[0, 1.0]$ for particle energies between $100 \, {\rm MeV}$ and $50\, {\rm MeV}$.
}
\begin{ruledtabular}
  \begin{tabular}{lccc}
    & {AMS in LEO} & {Geosynch.} & {Lagrange Pt.}  \\
    \hline
    $Q_{\rm min} \, [{\rm MeV}]$ &  500 & 50  & 5  \\
    $\tau$ & 90 minutes & 1 day & No rotation \\
    $a \, [{\rm m}^2{\rm sr}]$ & 0.055 & 0.165 & 0.165  \\
    $\theta_{\rm FOV} \, [{\rm rad}]$ & 0.168 & 0.292 & 0.292  \\
    $\tau_{\rm obs}(Q) $ & [0,1.0] & 1.0 & 1.0  \\

  \end{tabular}
\end{ruledtabular}
\end{table}

\subsubsection{Simulated PBH Signals}
\label{sec:PBHSignals}
We first compute the time-dependent positron signal generated by a transiting PBH and measured by an ideal detector in orbit around the Earth. The orbit, which approximates that of the ISS, is parameterized by a height $h=450$ km, a period $\tau=90$ min, and an inclination of $\pm 51.6^o$ latitude. Our ideal detector, which measures positrons with $100\%$ efficiency, has a field of view (FOV) angle $\theta_{\text{FOV}}$, an oriented area $\vec{A}(t)$, an energy-dependent acceptance $a(Q)$, and a minimum detectable positron energy $Q_{\rm min}$. This $Q_{\text{min}}$ parameter is determined in practice by the detector geometry, efficiency, trigger, and orbit relative to the geomagnetic field. For AMS, we take $Q_{\rm min} = 500 \, {\rm MeV}$, as discussed in Section \ref{sec:AMSDetector}. 

We use a 2D geometry. The PBH trajectory past Earth is parameterized by  $\vec{r}(t) = \bar{v} t\hat{x} + b\hat{y}$, with average velocity $\bar{v}= 246 \, {\rm  km/s}$ 
and impact parameter $b \gg R_{\oplus} + h$. These parameters are listed in Table \ref{table:SimPars} and a schematic of the transit geometry is shown in Fig.~\ref{fig:TransitGeometry}. The two independent parameters of this model are $M$ and $b$; velocity $v=\bar{v}\equiv246 \, {\rm km/s}$ is held constant. 
Parameters that depend on the detector configuration include: orbital period $\tau$, acceptance $a(Q)$, FOV angle $\theta_{\rm FOV}$, and minimum detectable positron energy $Q_{\rm min}$.

\begin{table*}
\caption{\label{table:SimPars}\justifying Symbol definitions and values for PBH transit simulation parameters and relevant variables. Only PBH mass $M$ and transit impact parameter $b$ are independent simulation parameters. The detector orbit and geometry parameters vary depending on which detector configuration is being considered, and are tabulated in Table~\ref{table:DetConfigs}. 
}
\begin{ruledtabular}
\begin{tabular}{llll}
Symbol & Definition   & Units \\ 
\hline
$M$         & PBH mass           & g \\
$b$         & Impact parameter   & AU  \\
$\bar{v}$         & Average PBH velocity      & km/s \\
$t$         & Time since closest approach & s\\
$\vec{r}(t)$&   PBH position     & AU\\
$Q$         & Positron energy  & GeV\\
$h$         & Orbit altitude       & km \\
$\tau$      & Orbit period       & minutes\\
$\theta_{\text{FOV}}$    & Detector field of view angle          & radians \\
$\vec{A}(t)$& Oriented detector area   & m$^2$\\
$a(Q)$      & Detector acceptance for $e^{\pm}$ with energy $Q$ & m$^{2}$sr\\      
$Q_{\text{min}}$   & Detector minimum detectable $e^{\pm}$ energy & GeV\\
$\tau_{\text{obs}}(Q)$& Observation time fraction for $e^{\pm}$ with energy $Q$ & \\
$\zeta_{e^+}(t)$ & Measured PBH positron count-rate signal & s$^{-1}$\\
$\mathcal{A}$ & Amplitide of measured PBH positron rate count-rate signal & \\
$\mu$       &lognormal model location parameter & g\\
$\sigma$    &lognormal model  width parameter & \\
$\alpha$       &GCC low-mass tail model parameter & \\
$\beta$       &GCC high-mass tail model parameter & \\
$\bar{\Phi}_{e^{\pm}}$& $e^+$-producing PBH transit rate past Earth & yr$^{-1}$\\
$b_{\text{max}}(M)$& Maximum impact parameter for detectable transit & AU\\
$\phi_i$ & Measured AMS $e^{\pm}$ flux for energy bin $i$ & m$^{-2}$s$^{-1}$sr$^{-1}$GeV$^{-1}$ \\
\end{tabular}
\end{ruledtabular}
\end{table*}

Given a PBH mass $M$, we numerically compute the secondary positron Hawking emission spectrum, $d^2N_{e^+}^{(2)}/dtdQ$,  using \texttt{BlackHawk v2.2}. Figs.~\ref{fig:SpectraComparison}--\ref{fig:PSComparison} show secondary positron spectra for several PBH masses in the positron-producing range. 

To determine the observable positron flux, we numerically integrate the secondary emission spectrum weighted by the energy-dependent observation time fraction $\tau_{\rm obs}(Q)$ between $Q_{\rm min}$ and some $Q_{\rm max}$. The observation time fraction is a dimensionless parameter that varies with particle energy $Q$ which represents the fraction of time that the detector is at a magnetic latitude with cutoff rigidity $R_c \leq Q$. We compute $\tau_{\rm obs}(Q)$ by interpolating the daily AMS energy-dependent observation times from Ref. \cite{aguilar_temporal_2023}, assuming that $\tau_{\rm obs}=0$ for $100 \, {\rm MeV}$, and dividing by the number of seconds in a day. Note that, in this context, ``observable'' refers to energies $Q\geq Q_{\rm min}$, which imposes that the emitted positions have an energy greater than the detector's minimum threshold. 

The upper bound cut at $Q_{\rm max}$ is more arbitrary, and should be chosen to capture the peak of the Hawking emission spectrum while avoiding background so as to maximize the signal-to-noise ratio (SNR). 
We take a simple approach here and define $Q_{\rm max}(M)$ to depend on the PBH mass such that it is larger than $Q_{\rm min}$ and satisfies: 
\begin{equation}
    \int_0^{Q_{\rm max}}\frac{d^2N^{(2)}}{dtdQ}dQ \bigg/ \int_0^{\infty}\frac{d^2N^{(2)}}{dtdQ}dQ \geq 0.99.
\end{equation}
This method of defining $Q_{\rm max}$ captures the greatest available signal but is not optimized to reduce noise and thereby maximize SNR. Improved methods would include making cuts to avoid the noisiest energy range around $\sim 1 \, {\rm GeV}$ and/or performing this analysis separately on many energy bins rather than integrating up the entire signal with respect to energy.

The total emission rate of observable positrons emitted by the transiting PBH of mass $M$ is thus:
\begin{equation}
    \label{eqn:ToyModelPBHFlux}
    \dot{N}(M) = \int_{Q_{\text{min}}}^{Q_{\rm max}} dQ \,\tau_{\rm obs}(Q) \frac{d^2N_{e^+}^{(2)}}{dtdQ}(M,Q).
\end{equation}

The weighting by $\tau_{\rm obs}$ in Eq.~(\ref{eqn:ToyModelPBHFlux}) approximately accounts for the changing geomagnetic cutoff due to variations in latitude during the orbits. Estimating observation time for positrons in a given energy bin is quite complicated for a real detector like AMS, which not only orbits the Earth and traverses $\pm 51.6^o$ of latitude, but is subject to various ISS maneuvers which change its pointing dynamically with time. Furthermore, the geomagnetic field is dynamic and has many features not captured by Stoermer's dipole approximation, such as the South-Atlantic Anomaly, where field intensity drops to about 1/3 of the typical intensity and particle fluxes increase dramatically \cite{pavon-carrasco_south_2016}. A more accurate determination of observation time as a function of positron energy would require ISS trajectory and altitude data, AMS orientation data (which is collected by an onboard GPS), and the International Geomagnetic Reference Field---a precise model of the time-varying geomagnetic field \cite{alken_international_2021}. This is beyond the scope of this analysis. 

We assume ballistic transport to propagate the emitted positrons to Earth and ignore the diffusive transport effects characteristic of charged particle propagation. This is a good approximation in our regime of interest because the parallel and perpendicular mean free paths for $e^{\pm}$ diffusion exceed $10^{-3}$ AU for all particle energies $Q\geq 5 \, {\rm MeV}$ at a distance of $1 \, {\rm AU}$ in the heliospheric magnetic field \cite{shalchi_nonlinear_2009, effenberger_generalized_2012, burger_fisk-parker_2008}. The mean free paths set the typical length scale that a charged particle will travel before scattering off a magnetic field inhomogeneity. 

We define the angle between the PBH at point $\vec{r}(t)$ and the detector area vector $\vec{A}(t)$ as:
\begin{equation}
    \label{eqn:PBHangle}
    \theta(t) \equiv \cos^{-1}{\left( \frac{\vec{r}(t) \cdot \vec{A}(t)}{|r(t)|| A(t)|} \right)},
\end{equation}
which is labeled in Fig.~\ref{fig:TransitGeometry}. 

\begin{figure}[t!]
\centering
\includegraphics[width=0.95\textwidth]{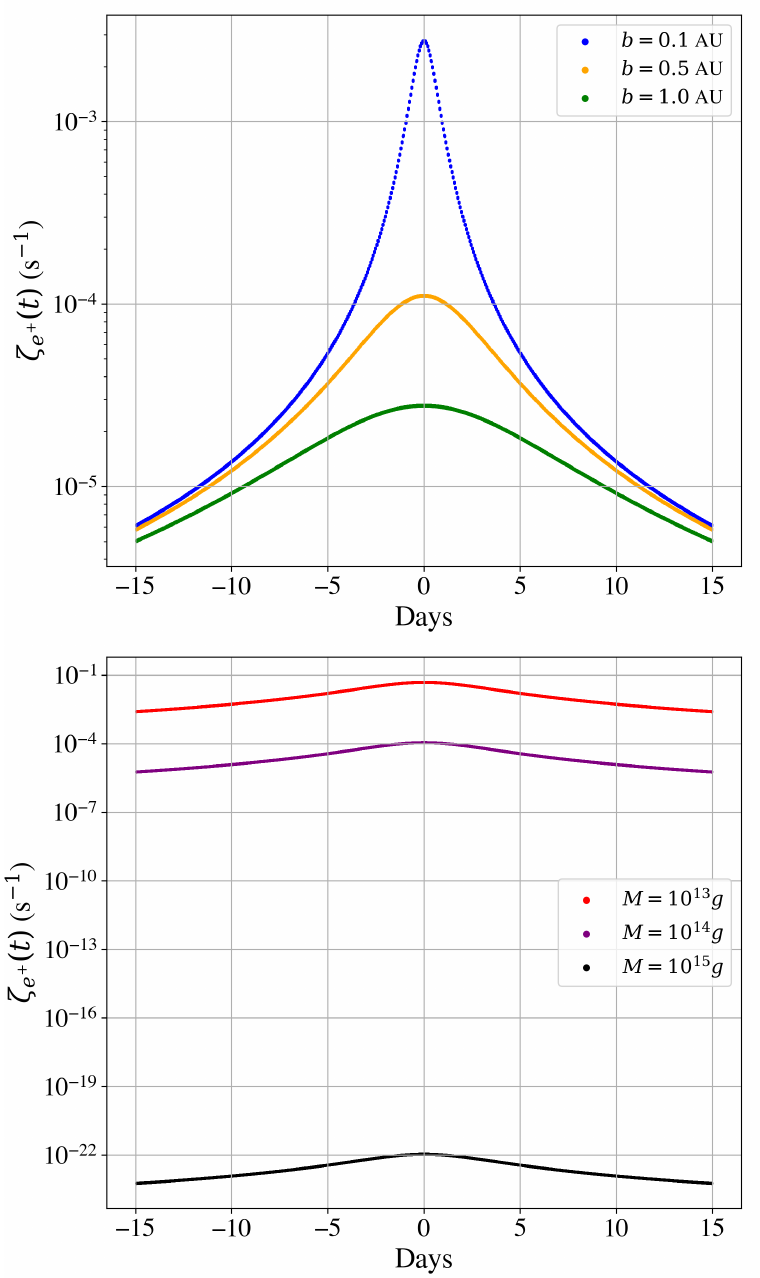}
\caption{\justifying Time-dependent positron count-rates $\zeta(t|M, b)$ for PBHs transiting past an ideal detector in orbit around Earth. These count-rates are calculated from Eq.~(\ref{eqn:ToyModelDetectorFlux}). Note that the curves appear to be dotted lines because they are highly oscillatory functions with a Lorentzian envelope. The oscillations have period $\tau \ll$ FWHM determined by the detector orbital frequency, and the count-rate signal is zero when the PBH is outside the detector FOV, causing the signal to vanish periodically. Time $t=0$ corresponds to the point of closest approach to Earth. The signal for a $10^{15} \, {\rm g}$ PBH (green curve, bottom panel) is negligible because AMS is only sensitive to positrons with energies $Q\geq 500 \, {\rm MeV}$ and the Hawking emission spectrum for $10^{15} \, {\rm g}$ PBHs, as shown in Fig.~\ref{fig:SpectraComparison}, cuts off around $100 \, {\rm  MeV}$. 
}
\label{fig:ToyModelSignals}
\end{figure}

We can then define the time-dependent positron signal count rate measured by our ideal orbiting detector as $\zeta_{e^+}(t)$, with dimensions $s^{-1}$:
\begin{equation}
\label{eqn:ToyModelDetectorFlux}
    \begin{split}
    \zeta_{e^+}(t|M, b) = &\frac{A \dot{N}(M)}{4 \pi (\bar{v}^2 t^2 + b^2)^{3/2}} \Theta( \theta_{\text{FOV}} - \theta(t)) \\ & \times \Biggl[ vt \sin{\left(\frac{2\pi}{\tau}t\right)}+ b \cos{\left(\frac{2\pi}{\tau}t\right)} \Biggr]. 
    \end{split}
\end{equation}
Note that this expression for $\zeta_{e^+}(t)$ quantifies the incident positron count rate in the energy range from $Q_{\text{min}}$ to $Q_{\text{max}}$ as measured by an orbiting detector; for a stationary detector, we only need the terms in the first line of Eq.~(\ref{eqn:ToyModelDetectorFlux}). The integrated Hawking emission rate $\dot{N}(M)$ is defined in Eq.~(\ref{eqn:ToyModelPBHFlux}). The Heaviside theta function $\Theta( \theta_{\text{FOV}} - \theta(t))$ imposes that the signal vanishes where $\theta(t) > \theta_{\text{FOV}}$, corresponding to intervals when the PBH is outside the FOV cone of the detector. Note that we use the assumption $b \gg R_{\oplus}+h$ to approximate the orbit of the detector about the Earth as rotation of the detector about a fixed point with period $\tau$. See Fig.~\ref{fig:ToyModelSignals} for plots of count-rate $\zeta_{e^+}(t)$ for the AMS experiment for various PBH masses and impact parameters. 

The FOV angle $\theta_{\rm FOV}$ that appears in Eq.~(\ref{eqn:ToyModelDetectorFlux}) can be computed from the detector acceptance and a few reasonable assumptions. Detector acceptance is the quantity that allows one to translate between a particle flux---a physical quantity independent of the measuring instrument---and the particle count rate measured by a specific instrument over a given time interval. We can define an energy-dependent acceptance as the product of the detector efficiency $\epsilon$, which includes contributions from trigger and selection efficiencies, and the detector geometric acceptance $a_{\rm geo}$:
\begin{equation}
    \label{eqn:acceptance}
    a(E) = \epsilon(E) \, a_{\rm geo}(E).
\end{equation}
The geometric acceptance, with dimensions ${\rm m}^2{\rm sr}$ is effectively the product of the detector's sensitive area $A$ and the solid angle $\Omega$ swept out by its field of view.
Assuming 100\% efficiency for our idealized detector ($\epsilon=1$), we can thus relate known quantities $a$ and $A$ to $\theta_{\text{FOV}}$ via the geometric acceptance:
\begin{equation}
    \label{eqn:thetaFOV}
    a = A\Omega = A \int_0^{2 \pi} d\phi \int_0^{\theta_{\text{FOV}}} d\theta \, \sin{\theta}.
\end{equation}
We take the detector sensitive area to be the size of one of the nine silicon tracker layers, which are not uniform in area. A representative area can be approximated by dividing the total reported sensitive detector area $6.75 \, {\rm m}^2$ by 9, which yields $A=6.75/9 = 0.75 \, {\rm m}^2$ \cite{alpat_internal_2010}. We use the more conservative area value $A=0.62 \, {\rm m}^2$.

For the unmodified AMS detector configuration, we use the reported geometric acceptance $a_{\rm geo}^{\rm AMS}=0.055 \, {\rm m}^2{\rm sr}$ \cite{ams_collaboration_electron_2014}. For the two hypothetical detector scenarios, we note that the planned Layer-0 AMS upgrade predicts a $300\%$ increase in acceptance by the inclusion of two additional large-area silicon tracker layers to the top of the instrument stack \cite{ubaldi_charge_2023}. We thus optimistically take $a_{\rm geo}=3a_{\rm geo}^{\rm AMS} = 0.165$ for these cases. The derived values for $\theta_{\rm FOV}$ are thus $\theta_{\rm FOV}=0.153\, {\rm rad}$ for AMS and $\theta_{\rm FOV}=0.265\, {\rm rad}$ for the hypothetical large-acceptance configurations. 

\begin{figure}[t]
\centering
\includegraphics[width=0.95\textwidth]{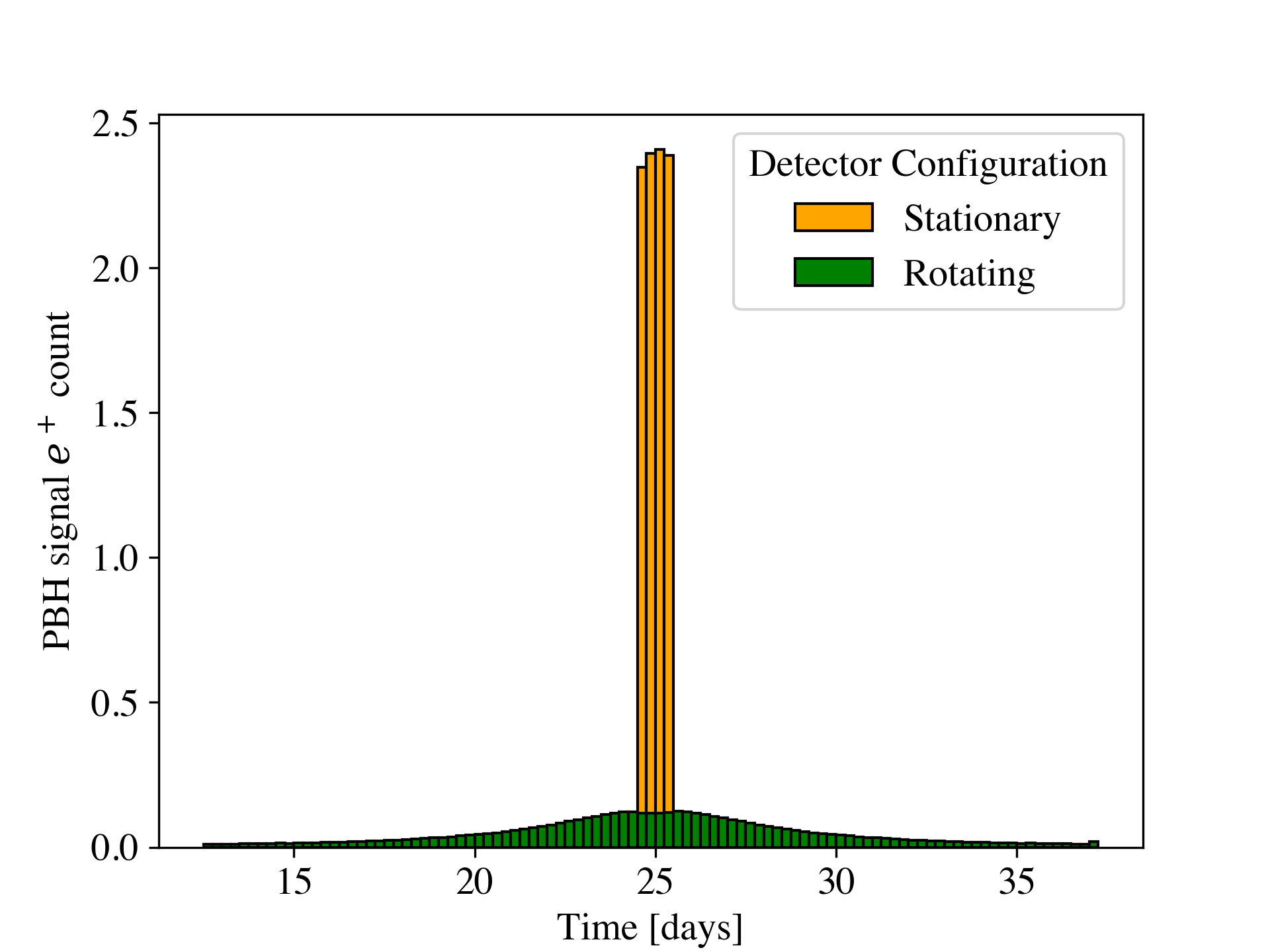}
\caption{\justifying Positron counts binned every 0.25 days for the measured signal generated by a $10^{14} \, {\rm g}$ PBH transiting past Earth with an impact parameter of $b = 0.5 \, {\rm AU}$. The signal observed by a stationary detector (yellow) is compared with the signal observed by a detector rotating with a period of 90 minutes (green). The integrated positron counts observed throughout the entire transit duration are 9.5 counts for the stationary detector and 5.1 counts for the rotating detector. The rotating signal duration is defined by a FWHM of 7 days. 
}
\label{fig:HistoSignal}
\end{figure}

Figure \ref{fig:HistoSignal} shows in green the positron signal count measured by our ideal AMS detector with no background for $M=10^{14}$ g and $b = 0.5 \, {\rm AU}$ binned over intervals of 0.25 days. 

The signal counts are binned over intervals of 0.25 days, thus averaging over 4 orbital cycles for each bin. Each time-dependent transit signal is characterized by its amplitude, time of closest approach, and full width at half maximum (FWHM), which can be used as a measure of the effective ``duration'' of the transit. For the Lorentzian envelope in Eq.~(\ref{eqn:ToyModelDetectorFlux}), the FWHM is determined by the PBH velocity and impact parameter:
\begin{equation}
    \label{eqn:FWHM}
    {\rm FWHM} = \frac{b}{v}\sqrt{2^{2/3}-1}.
\end{equation}
Figure \ref{fig:HistoSignal} compares the binned positron count signals measured by a stationary and rotating AMS detector. At time-of-closest-approach, the stationary signal is about 10 times larger, but its duration is quite short (1 day) due to the detector's narrow FOV, compared with the 7 day FWHM of the signal seen by the rotating detector. The resulting integrated positron counts differ by only a factor of 2 instead of 10, as discussed in the caption. Thus, implementing a stationary detector for the third detector configuration gives an SNR boost by a factor of $\sim 10$, but PBH detection is limited by the time-resolution of the binned data.  

To determine whether these time-dependent positron fluctuations are detectable with AMS, we must accurately simulate the time-varying positron background. Using a matched filter and proper binning in time, we find that we can detect low-mass PBH transits with SNR $\lesssim1$. 

\subsubsection{\label{sec:PositronBkgd}Positron Background}

Positrons are considered \textit{secondary cosmic rays}, defined as originating from the collisions of other cosmic rays with the interstellar medium (ISM). The dominant production mode for cosmic ray positrons is the inelastic scattering of high energy cosmic ray protons with protons in the ISM \cite{moskalenko_production_1998}. Positrons (and secondary electrons) are the final stable products of 
hadronic decays resulting from these high energy $pp$ collisions. Cosmic ray propagation models predict a power-law positron energy spectrum with $E^3\Phi$ peaked around $5$ GeV ($\Phi$ being the positron flux), which is in agreement with low-energy data from experiments like AMS, PAMELA, and Fermi-LAT \cite{aguilar_towards_2019}. The high-energy positron excess observed by these experiments \cite{adriani_anomalous_2009, aguilar_towards_2019, the_fermi-lat_collaboration_cosmic-ray_2017} is theorized to be of \textit{primary} origin from either dark matter annihilation \cite{jin_astrophysical_2020} or pair-production in pulsar magnetic fields \cite{hooper_pulsars_2009}. 

\begin{figure}[t]
\centering
\includegraphics[width=0.95\textwidth]{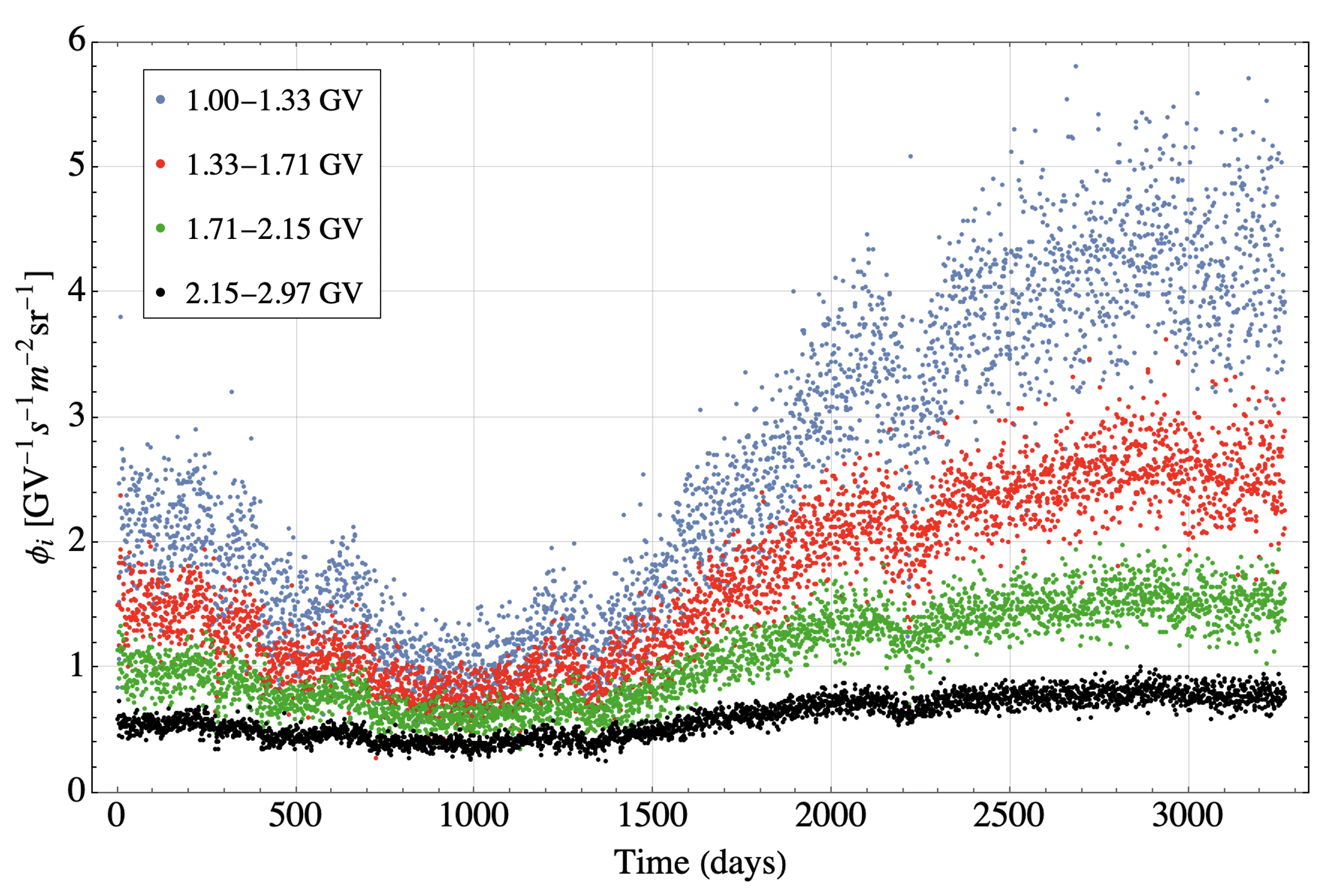}
\caption{\justifying AMS positron daily fluxes collected over approximately 9 years of observation. 
The data displays a period of 11 years, corresponding to the 11 year solar cycle. Solar modulation effects are most prominent for low energy positrons. We plot the four reported lowest energy bins.}
\label{fig:AMSdata}
\end{figure}

We are only concerned with the well-measured low-energy positron background below $\mathcal{O}(10 \, {\rm GeV})$, because positrons with $Q \gtrsim 10 \, {\rm GeV}$ are only produced in abundance by PBHs with masses below $10^{13} \, {\rm g}$. Temporal fluctuations of the $(e^- + e^+)$ cosmic ray spectrum were shown to vanish sharply when Voyager 1 crossed the heliopause and left the Solar System in 2012 \cite{rankin_galactic_2022}; however, Voyager still measures time-dependent, anisotropic features in the flux after episodes of solar activity \cite{rankin_galactic_2019}. We infer that the dominant source of background for temporal positron fluctuations---both periodic and irregular---is solar activity. Low-energy secondary positrons have well-measured and energy-dependent temporal fluctuations within the Solar System. (See Fig. \ref{fig:AMSdata}.) Large-amplitude fluctuations track the the 11-year solar cycle and small-amplitude fluctuations have been reported with periods matching the 27-day Bartels' rotation of the Sun relative to the Earth \cite{aguilar_temporal_2023}.  
The time-scales of the longest PBH transits considered here are on the order of weeks, implying that periodic fluctuations on the order of months or years can be neglected when simulating time-dependent background for such transits. 

We therefore focus on sampling the observed energy-dependent Gaussian noise to construct the positron background. 
We account for the fact that the mean flux and standard deviation of the flux for the underlying Gaussian noise are both energy-dependent and time-dependent (dominated by the 11-year solar cycle). See Fig. \ref{fig:AMSdata}. The mean fluxes and standard deviations for a given energy bin are lowest during the period of ``solar minimum'' ($t\sim1000 \, {\rm days}$) and largest during ``solar maximum'' ($t\sim4000 \, {\rm days}$). This modulation is strongest for positrons with $Q\lesssim 1 \, {\rm GeV}$, and approximately negligible for $Q\gtrsim 10 \, {\rm GeV}$. The procedure to simulate the time-dependent positron background using published AMS integrated positron flux data for energies $500 \, {\rm MeV} \leq Q \leq 1 \, {\rm TeV}$ and temporal daily flux data for energies $1 \, {\rm GeV} \leq Q \leq 41.9 \, {\rm GeV}$ from Refs.~\cite{aguilar_temporal_2023, aguilar_towards_2019} is described below.

For each PBH transit, positron fluxes at different energies are sampled from distinct Gaussian distributions, converted to particle counts, and summed together to form an energy-integrated background for the energy range $[Q_{\rm min}, \, Q_{\rm max}]$. For the $i$th energy bin of width $\Delta E_i$ and characteristic energy $\tilde{E}_i$, we calculate the mean positron background count rate as measured by our detector, $\langle  \zeta_{\text{bkgd}, i} \rangle$ with units ${\rm s}^{-1}$:
\begin{equation}
    \label{eqn:PositronBkgdRate}
    \langle \zeta_{\text{bkgd}, i} \rangle= \phi_i \,a(\tilde{E}_i) \,\Delta E_i \,\tau_{\rm obs}(\tilde{E}_i).
\end{equation}
Here $\phi_i$ is the flux for the $i$th bin in ${\rm m^{-2} \,sr^{-1} \, GeV^{-1} \, s^{-1} }$ as reported by the AMS Collaboration in Ref.~\cite{aguilar_towards_2019}, $a$ is the detector acceptance, and $\tau_{\rm obs}$ is the observation time fraction discussed in the previous section to account for the geomagnetic cutoff. (The flux $\phi_i$ should not be confused with the PBH number distribution $\phi (M)$.)

Whereas the mean of the Gaussian noise distribution for each bin $\langle  \zeta_{\text{bkgd}, i} \rangle$ can be computed via Eq.~(\ref{eqn:PositronBkgdRate}), the most important quantities to model are the standard deviations $\sigma_i$ of the fluxes for each bin. 
The SNR of a PBH transit signal is defined as 
\begin{equation}
    \label{eqn:SNR}
    \text{SNR} \equiv \frac{\mathcal{A}_{\text{max}}^2}{\sigma_{n}^2},
\end{equation}
where $\mathcal{A}_{\text{max}}$ is the signal amplitude and $\sigma_n$ is the standard deviation of the total energy-integrated temporal noise in the simulated background. 
Because the total noise is the sum of normally distributed random variables, we expect that $\sigma_n$ takes the form
\begin{equation}
    \sigma_n = \sqrt{\sum_i \sigma_i^2} \, .
\end{equation}
Therefore the SNR, and thereby the PBH detection efficiency, is determined by the variances of all the bins enclosed in the interval $[Q_{\rm min}, Q_{\rm max}]$. 

For each energy bin $i$, we use 100-day running averages to compute the average $\mu_i$ and the standard deviation $\sigma_i$ 
as functions of time, and determine the minimum and maximum values of $\sigma_i / \mu_i$ for each energy bin. The global minimum and maximum values of this ratio over all 12 energy bins are $(\sigma/\mu)_{\rm min} = 0.063$ and $(\sigma/\mu)_{\rm max} = 0.263$. Thus, for some energy bin $i$, a conservative estimate for the standard deviation of the fluxes at solar minimum and solar maximum are $\sigma_{i,{\rm min} } = \langle  \zeta_{\text{bkgd}, i} \rangle(\sigma/\mu)_{\rm min}$ and $\sigma_{i,{\rm max} } = \langle  \zeta_{\text{bkgd}, i} \rangle(\sigma/\mu)_{\rm max}$, respectively. Monte Carlo simulations assuming the ``noise ratio'' or 0.63 or 0.263 form the upper and lower bounds on the shaded regions in Fig.~\ref{fig:bMaxM} and the black lines are computed assuming the central value 0.163. 

\begin{figure*}[t]
    \centering
    \includegraphics[width=0.925\textwidth]{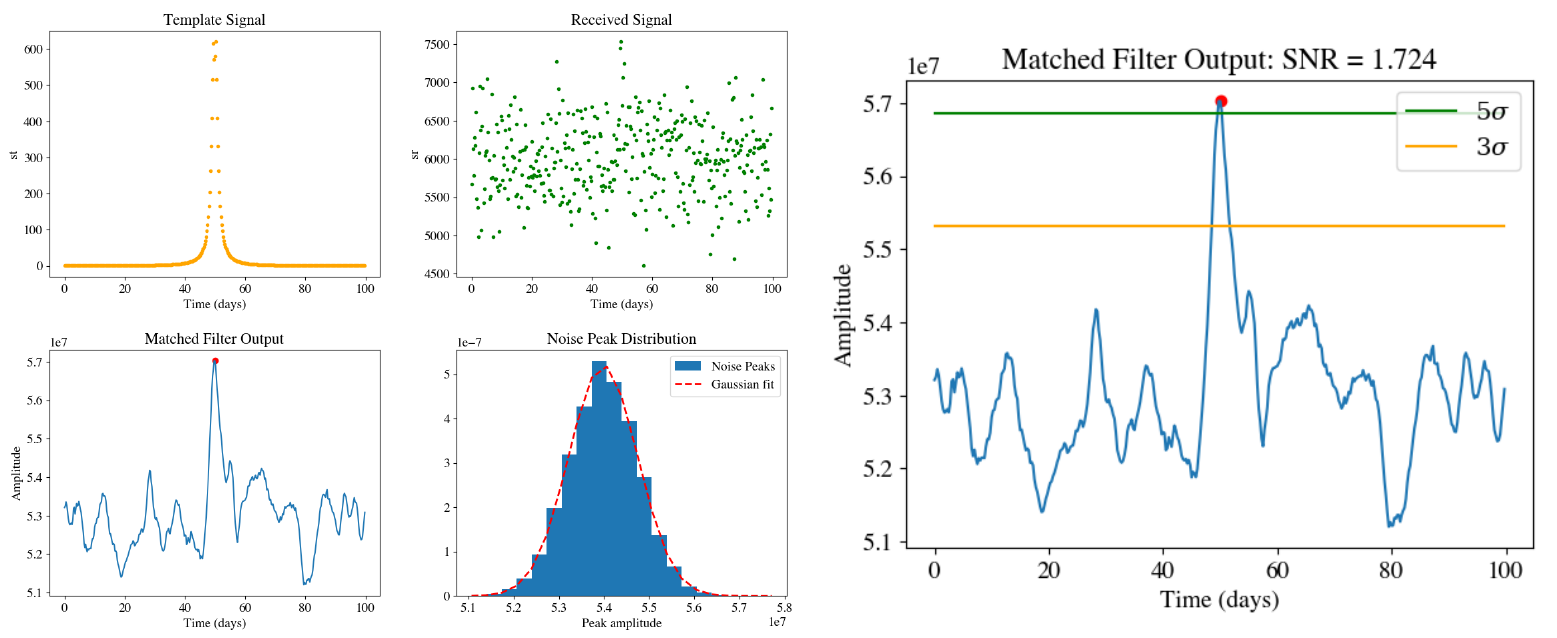}
    \caption{\justifying An example showing the intermediate steps for determining the time of closest approach and statistical significance of the matched filter output for a PBH transit event. This event had parameters $M (t_0) =5.2\times10^{13}  \, {\rm g}$ and $b=0.15 \, {\rm AU}$ and was considered a positive detection with significance above the $5\sigma$ threshold. The first panel (orange) shows the template signal counts binned on intervals of $0.25$ days. The second panel (green) shows the net received signal: the sum of the signal and noise. The histogram plots the peak amplitude distribution obtained by filtering $10^4$ days of noise. A gaussian fit is used to compute the noise peak standard deviation $\sigma$; a PBH event detection is considered significant if its amplitude exceeds $5\sigma$. A fit to the peak of the matched filter output (right panel, red point) identifies the estimated time of closest approach $t_0$. Horizontal lines indicate $3\sigma$ and $5\sigma$ thresholds.}
    \label{fig:SampleTransit}
\end{figure*}

To generate a daily flux background for each energy bin, we therefore sample from a Gaussian with mean $\langle \zeta_{\text{bkgd}, i} \rangle$, calculated according to Eq.~(\ref{eqn:PositronBkgdRate}), and standard deviation $\sigma_i$. We then convert the fluxes to counts and sum over all energy bins in the range $[Q_{\rm min}, Q_{\rm max}]$. An example of a sampled background superposed with the simulated PBH signal is shown in Fig.~\ref{fig:SampleTransit} (green points).

\subsubsection{\label{sec:MatchedFilter}PBH Detection}

Given that we can generate a PBH $e^+$ signal according to Sec.~\ref{sec:PBHSignals} and sample the time-series background according to Sec.~\ref{sec:PositronBkgd}, we now develop a matched filter technique to detect low-SNR PBH transit signals and estimate time of closest approach. We estimate the theoretical maximum detection efficiency for PBH transits by a detector configuration 
by running the matched filter with a known template signal. This allows us to compute maximum detectable impact parameter $b_{\text{max}}(M)$ for the detector and therefore to estimate the annual number of detectable PBH transits for any given PBH mass function model. 

A \textit{matched filter} is defined as the optimal linear filter to maximize SNR, as defined in Eq, ~(\ref{eqn:SNR}).
The input to the filter is the received signal:
\begin{equation}
    \label{eqn:MF_input}
    s_r(t) = s_t(t-t_0) + n(t),
\end{equation}
where $s_t(t-t_0) \equiv \zeta_{e^+}(t-t_0)$ is the PBH signal defined in Eq.~(\ref{eqn:ToyModelDetectorFlux}) peaked at time of closest approach $t_0$ and is referred to as the \textit{template signal}. The time-dependent noise background $n(t)$ is generated by the sampling process discussed in Sec.~\ref{sec:PositronBkgd}. The matched filter ${\cal H}(t)$ is defined as the conjugate of the scaled, time-reversed and time-shifted template signal, 
\begin{equation}
    \label{eqn:MF}
    {\cal H}(t) = K s_t^*(t_0-t) ,
\end{equation}
where the scaling constant $K$ is typically set to 1.

The output of the matched filter, $s_o(t)$, is the convolution of ${\cal H}(t)$ and $s_r(t)$:
\begin{equation}
    \label{eqn:MF_output}
    s_o(t) = \mathcal{F}^{-1}\{\mathcal{F}\{ {\cal H}(t)\}\times\mathcal{F}\{s_r(t)\}\},
\end{equation}
where $\mathcal{F}$ denotes the Fourier transform.

To determine whether a PBH transit signature is present in a simulated received signal dataset $s_r(t)$ and to estimate the time of closest approach, we first construct the matched filter ${\cal H}(t)$ given the transit parameters $M, b, \text{ and } v$. We apply the filter by convolving $s_r(t)$ with $\mathcal{H}(t)$ and perform peak detection on the filtered output $s_o(t)$ to estimate the amplitude and location of the largest peak. We then perform a numerical Monte Carlo study to simulate the peak amplitude distribution generated by running $\mathcal{O}(10^4)$ days of simulated noise through the filter ${\cal H}(t)$. The detected peak in the filtered output $s_o(t)$ is compared to this distribution of amplitudes generated by the noise Monte Carlo study and accepted as a positive detection if its amplitude exceeds a $5\sigma$ threshold. Figure \ref{fig:SampleTransit} shows the results of a matched filter analysis and significance study for a simulated PBH transit with parameters $M=5.2\times10^{13} \, {\rm g}$, $b=0.15 \,  {\rm AU}$, and $v=246 \,  {\rm km/s}$ and with the temporal counts binned every $0.25$ days. This example has an SNR of $1.74$ and is detected at a significance level above $5\sigma$ using a known template signal.

\subsubsection{\label{sec:Detectability}Maximum Impact Parameters}

The goal of developing this matched filter signal detection technique is to run numerical studies to estimate an upper bound on the PBH detection efficiency of an idealized orbiting detector for a given set of transit parameters $(M, b)$. Note that we define detection efficiency to be the fraction of PBHs with transit parameters $(M, b)$ which are detected above the $5\sigma$ level by our detector. Our results are an upper bound on detection efficiency because the matched filter uses a template signal where parameters $M, b, \text{ and } v$ are known. 

\begin{figure}[t]
\centering
\includegraphics[width=0.95\textwidth]{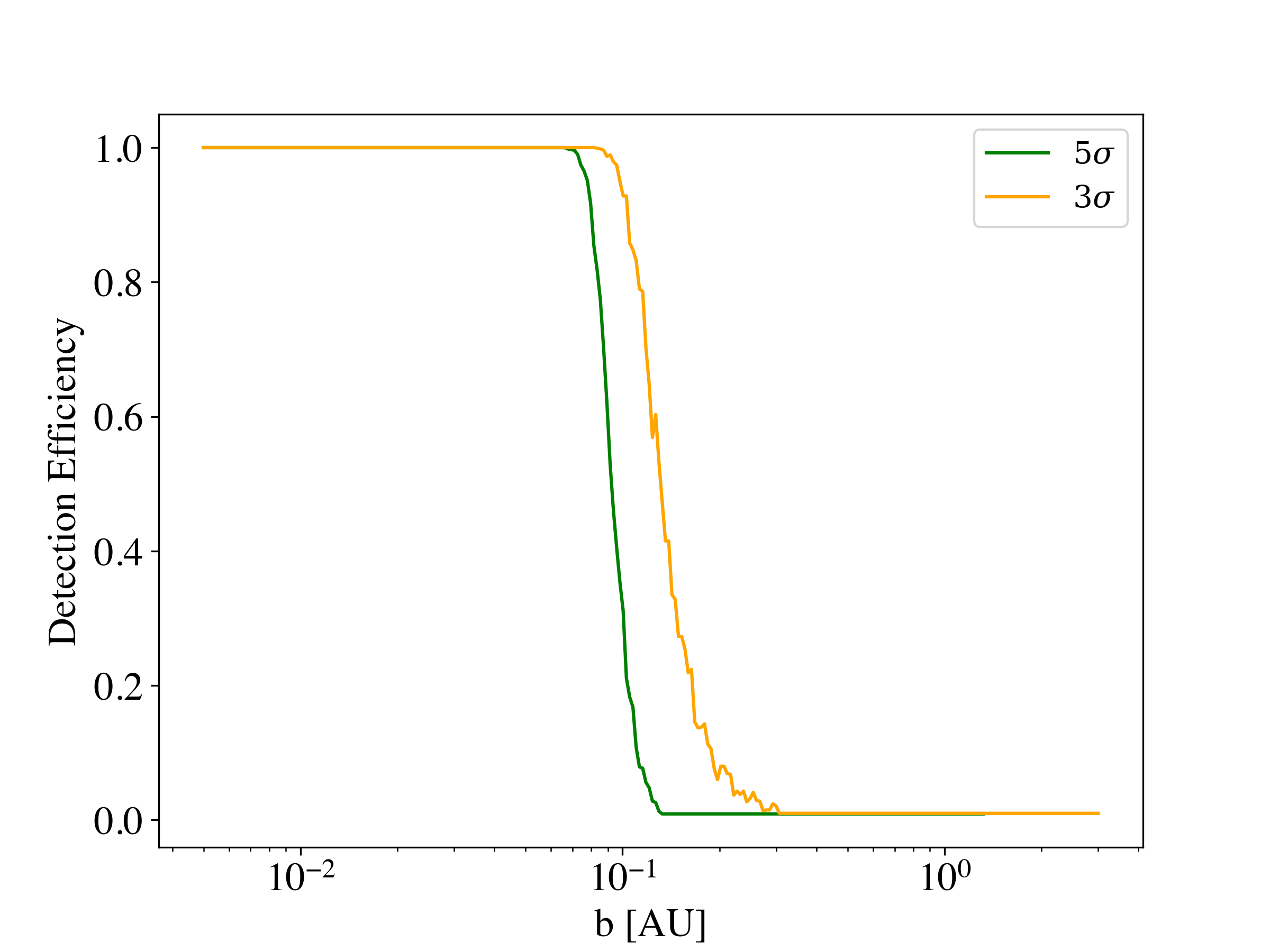}
\caption{\justifying PBH detection efficiency as a function of impact parameter $b$ for PBHs with mass $M (t_0) =10^{14} \, {\rm g}$ assuming the AMS detector configuration. Detection efficiency is computed numerically via Monte Carlo simulations for two different significance levels, $3\sigma$ and $5\sigma$. The maximum impact parameter $b_{\rm max}=(6.68\pm0.07)\times10^{-2} \, {\rm AU}$ is defined as the value of $b$ for which the detection efficiency is 99\% for the $5\sigma$ curve. }
\label{fig:DetEff}
\end{figure}

Given that we can use numerical studies to determine detection efficiency, as plotted for example parameters $M=10^{14} \, {\rm g}, \, b=0.15 \, {\rm AU}$ in Fig.~\ref{fig:DetEff}, we may next determine a \textit{maximum impact parameter} function, $b_{\text{max}}(M)$, for each detector configuration. We define $b_{\rm max}$ as the largest impact parameter such that a PBH with mass $M (t_0)$ can be detected by a given detector geometry at $5\sigma$ significance with 99\% efficiency. The general steps followed to calculate $b_{\text{max}}$ for a specified value of $M (t_0)$ are: (1) simulate $\mathcal{O}(10^4)$ days of background, (2) for a given value of $b$, generate the optimal filter $ {\cal H}(t|M, b, v)$, (3) run the noise through the filter ${\cal H}(t)$ to generate a background peak amplitude distribution, (4) simulate 1000 PBH transits with the given $b$ value and determine filtered peak amplitude significance for each one, (5) compute detection efficiency (the fraction of the 1000 transits which are detected at or above the $5\sigma$ level), (6) repeat steps 2-4 for impact parameters in the range $[10^{-3}\,  {\rm AU}, 10^2\,  {\rm AU}]$, (7) identify $b_{\text{max}}(M)$ as the value of $b$ at which detection efficiency drops below 99\%. Figure \ref{fig:DetEff} plots detection efficiency as a function of impact parameter $b$ for PBH mass $M (t_0)=10^{14} \, {\rm g}$ using two different significance thresholds and assuming the AMS detector configuration.

\begin{figure*}[t]
\centering
\includegraphics[width=0.7\textwidth]{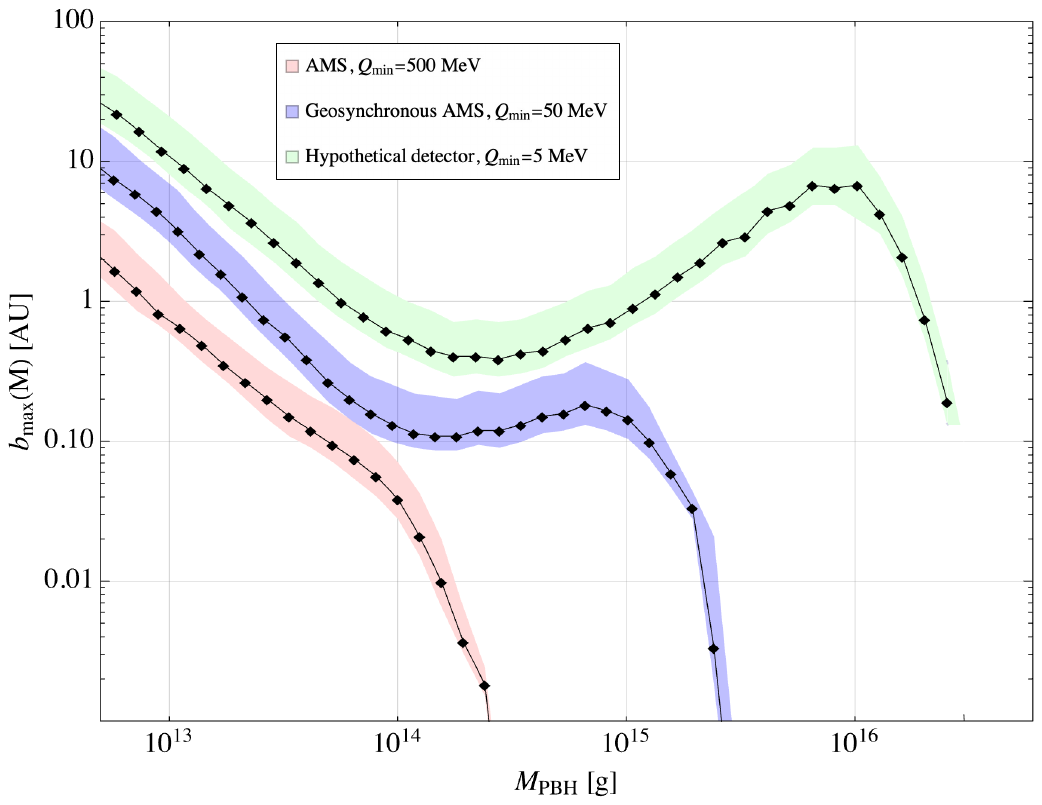}
\caption{\justifying Maximum impact parameter as a function of PBH mass $M (t_0)$ for three datasets: the AMS experiment with $Q_{\text{min}}=500 \,  {\rm MeV}$ (red), AMS in geosynchronous orbit with $Q_{\rm min} = 50 \, {\rm MeV}$ (blue), and a hypothetical detector with AMS acceptance unchanged but $Q_{\rm min}=5 \, {\rm MeV}$ and no geomagnetic field (green). Data points in this figure are results of the numerical studies described in Sec.~\ref{sec:Detectability}. Lines are to guide the eye. AMS would only be sensitive to PBHs with masses $M (t_0) \lesssim10^{14} \, {\rm g}$ due to its relatively high lower threshold energy for detecting positrons. Note that the Earth-Moon distance is about $0.0025  \, {\rm AU}$. The lower and upper bounds on the shaded regions correspond to assuming either solar maximum or solar minimum conditions, respectively, when simulating the time-dependent positron background. Black lines and points are computed assuming the central value for the noise fraction between solar maximum and minimum. 
}
\label{fig:bMaxM}
\end{figure*}

Figure \ref{fig:bMaxM} plots $b_{\text{max}}(M)$ for masses $5\times10^{12} \, {\rm g} \leq M (t_0) \leq 10^{17} \, {\rm  g}$ for the three detector configurations. For the AMS detector configuration (red), the maximum impact parameter drops below $10^{-2} \, {\rm  AU}$ for $M (t_0)\approx 2\times 10^{14} \,  {\rm g}$. The predominant limitation that prevents AMS from being sensitive to transiting PBHs with masses $ M >M_*$ is the energy range of AMS positron data. Because $Q_{\text{min}}=500\,{\rm MeV}$, AMS will only be sensitive to PBHs with secondary positron spectra that peak above $500 \,  {\rm MeV}$. Secondary emission spectra typically peak at an energy $Q_{\text{peak}} \approx 5T$ for a PBH of temperature $T$. Thus, AMS should only be sensitive to PBHs with temperatures $T\gtrsim100 \,  {\rm MeV}$, which corresponds to $M (t_0)\lesssim10^{14} \, {\rm g}$, which is reflected in our result. 

The maximum impact parameter curves in Fig.~\ref{fig:bMaxM} for the $Q_{\rm min}=50 \, {\rm MeV}$ (blue) and $Q_{\rm min}=5 \, {\rm MeV}$ (green) datasets show more complex behavior and admit detectable PBHs up to $\sim3\times10^{15} \, {\rm g}$ and $\sim3\times10^{16} \, {\rm g}$, respectively. Both curves feature a local minimum around $10^{14} \, {\rm g}$. This is due to the energy-dependent noise peaking around $1 \, {\rm GeV}$---about the temperature of a $10^{14} \, {\rm g}$ black hole. PBHs with secondary positron spectra that peak near $1 \, {\rm GeV}$ thus have noisier background than those peaked at lower energies. Furthermore, positron Hawking spectra that peak around $T=1 \, {\rm GeV}$ have lower integrated emission rates than PBHs with temperatures $T\gg1 \, {\rm GeV}$ due to the decrease of $\dot{N}(M)$ with energy. These features of the Hawking spectra and positron background result in a local minimum around $M=10^{14} \, {\rm g}$. The subsequent turnover of the curves observed at $\sim10^{15} \, {\rm g}$ and $\sim10^{16} \, {\rm g}$ (respectively) corresponds to the positron Hawking spectrum peak occurring at energies $Q<Q_{\rm min}$ for each dataset, thus resulting in low signal counts.

\subsubsection{\label{sec:ModelsProbed}Sensitivity to PBH Models}

Using the values of $b_{\text{max}}(M)$ shown in Fig.~\ref{fig:bMaxM} from our numerical studies, we can produce an updated version of Fig.~\ref{fig:TransitRate}, which now plots an upper bound on the number of \textit{detectable} $e^{\pm}$-producing PBH transits per year for each detector configuration as a function of LN and GCC model parameters. We compute the detectable PBH transit rate by replacing the $b_{\text{max}}$ integration bound of Eq.~(\ref{eqn:ImpactParamIntegral}) with the mass dependent $b_{\text{max}}(M)$. The number of detectable transits per year is then 
\begin{equation}
    \label{eqn:DetTransits}
    \bar{\Phi}_{e^{\pm}}^{\rm det.} = \rho_{\rm DM}^{\odot}\int_{0}^{5\times10^{17} {\rm g}} \frac{dM}{M} \psi(M, t_0) \bar{v} \pi b_{\rm max}^2(M),
\end{equation}
which is integrated numerically by interpolating the $b_{\max}(M)$ values shown as black points in Fig. \ref{fig:bMaxM} and using the LN and GCC mass functions $\psi(M, t_0)$ defined in Section \ref{sec:PBHMassFunctions}.

\begin{figure*}[t]
\centering
\includegraphics[width=1.0\textwidth]{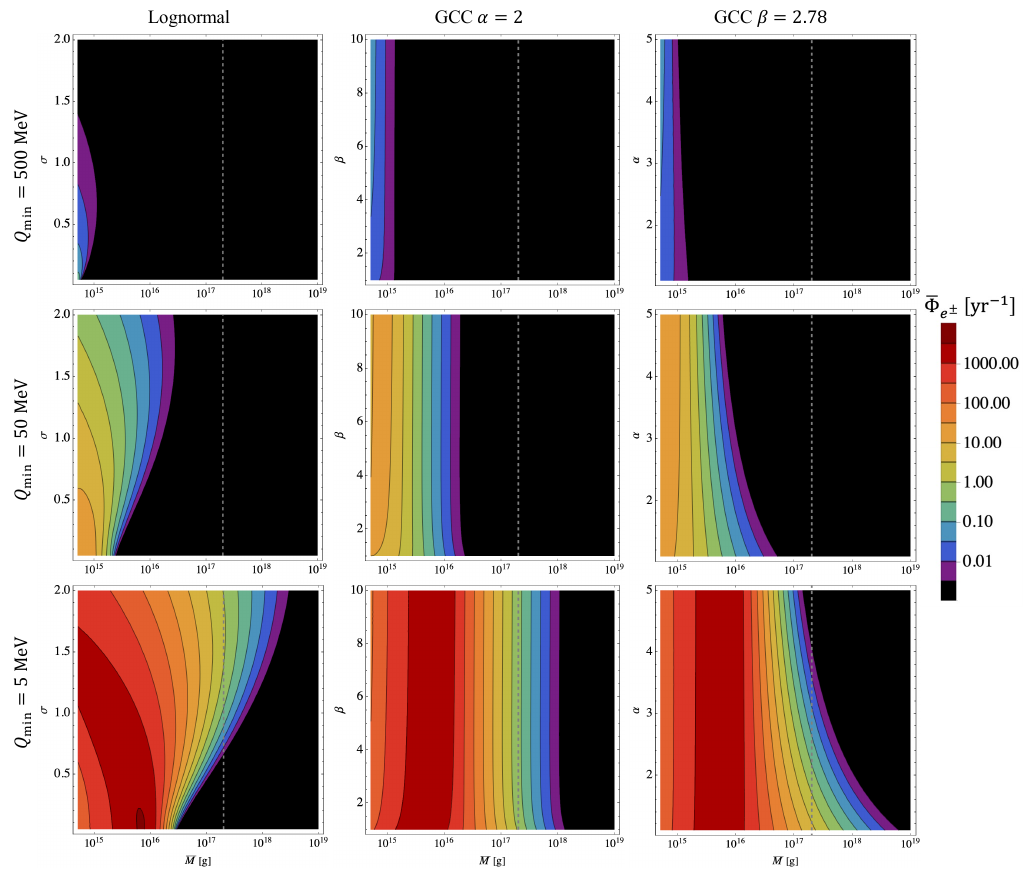}
\caption{\justifying Total number of detectable $e^{\pm}$-producing PBH transits per year as a function of $\bar{M}_i$ at PBH formation time for three simulated datasets: AMS, which is sensitive to positrons with energies $Q \geq Q_{\rm min} = 500 \, {\rm MeV}$ (top row), AMS in geosynchronous orbit, sensitive to positrons with $Q \geq Q_{\rm min} = 50 \, {\rm MeV}$ (middle row), and a non-rotating AMS-like detector  beyond the geomagnetic field with $Q_{\rm min}=5 \, {\rm MeV}$. Note that we define a ``detectable'' transit to be a PBH of mass $M (t_0)$ transiting with impact parameter $b\leq b_{\rm max}(M)$, as plotted in Fig. \ref{fig:bMaxM}. For all models with $\bar{M}_i\geq2\times10^{17} \, {\rm g}$ (vertical dashed gray line), $f_{\rm PBH}=1$ is allowed by existing Hawking emission constraints on extended mass functions \cite{luque_refining_2024,gorton_how_2024}. }
\label{fig:DetTransits}
\end{figure*}

Figure~\ref{fig:DetTransits} plots $\bar{\Phi}_{e^{\pm}}^{\rm det.}$ for regions of LN and GCC parameter with $\bar{M}_i>M_*$ for all three detector configurations, labeled by their respective $Q_{\rm min}$ values. With 10 years of time-dependent positron flux data, the AMS experiment could constrain very small regions of both extended mass function model parameter spaces. The main difficulty that prevents AMS from probing a larger region of parameter space is its high value of $Q_{\rm min}=500 \, {\rm  MeV}$, which limits its sensitivity to PBHs with temperatures $T\gtrsim100 \, {\rm MeV}$ or, equivalently, masses $M\lesssim 10^{14} \, {\rm g}$. The hypothetical $Q_{\rm min}=50 \, {\rm MeV}$ and $Q_{\rm min}=5 \, {\rm MeV}$ positron datasets (rows 2 and 3 of Fig.~\ref{fig:DetTransits}) would be capable of probing much larger regions of parameter space.

For the geosynchronous detector configuration, we re-compute $b_{\text{max}}(M)$ assuming all AMS detector parameters held constant except $Q_{\text{min}}=50 \, {\rm MeV}$, $\tau_{\text{obs}}=1$ for all $Q$, and orbital period $\tau=1$ day. (See the blue curve in Fig.~\ref{fig:bMaxM}.) We then use this mass-dependent integration bound in Eq.~(\ref{eqn:ImpactParamIntegral}) to compute the number of detectable transits per year given a set of mass function model parameters. The results for this hypothetical dataset are shown as the middle row of contour plots in Fig.~\ref{fig:DetTransits}. 

Finally, we repeat the analysis for a hypothetical detector that has equivalent precision and acceptance to AMS but is fully outside the Earth's magnetosphere, and hence would be sensitive to positrons with energies $Q \geq Q_{\rm min} = 5 \, {\rm MeV}$. Contour plots of detectable transit rates $\bar{\Phi}_{e^{\pm}}^{\rm det.}$ for this detector configuration are shown in the bottom row of Fig.~\ref{fig:DetTransits}. This simulated dataset is capable of probing PBH mass functions peaked in the asteroid mass range ($\bar{M}_i\gtrsim2\times10^{17} \, {\rm g}$) even though maximum impact parameters only exceed $10^{-3} \, {\rm AU}$ for $M\leq3\times10^{16} \, {\rm g}$ (see the green curve in Fig.~\ref{fig:bMaxM}), because extended mass functions have a low-mass tail which yields a nontrivial subpopulation of PBHs with mass $M\leq\bar{M}_i$.

\section{\label{sec:Conclusion}Conclusion}
In order to close the unconstrained asteroid-mass window in PBH mass parameter space, new detection signatures are required.  We demonstrate a novel method of analyzing time-series cosmic ray positron data to look for evidence of PBHs transiting past Earth through the inner Solar System.  Positrons are a particularly promising signature due to their low astrophysical backgrounds and the high precision and long observation time period of AMS data. We have shown that if PBHs obey a realistic extended mass function $\psi (M_i)$ that is peaked at $\bar{M}_i \geq 2 \times 10^{17} \,{\rm g}$, for which $f_{\rm PBH} = 1$ remains viable in the light of existing observations and constraints \cite{escriva_primordial_2024,Carr:2009jm, carr_primordial_2020,green_primordial_2021,carr_constraints_2021,carr_primordial_2022,Ozsoy:2023ryl,carr_observational_2024,gorton_how_2024} , then we should expect ${\cal O} (3)$ or fewer transits per year to pass within $1 \, {\rm AU}$ of the Earth. These high transit rates motivate us to simulate the time-dependent positron signature measured by an experiment like AMS in LEO during a PBH transit. 

We find that AMS would not be sensitive to PBHs with present-day masses $M (t_0) \gtrsim2\times10^{14} \, {\rm g}$ because of its sensitivity to positron energies $Q \geq Q_{\rm min} = 500 \, {\rm MeV}$. We have demonstrated that this new technique may be effective, however, if applied to a dataset that reaches down to low enough positron energies to probe the larger-mass PBHs that may exist today. In order to probe PBH mass functions that peak below $\bar{M}_i \leq 5 \times 10^{17} \, {\rm g}$, we would need time-dependent positron flux data from at least several years of observations spanning energies from approximately $5 \, {\rm MeV} - 1\, {\rm GeV}$. To collect such low-rigidity positrons would require that the instrument be located outside the Earth's magnetosphere. If a positron detector with similar precision and acceptance as AMS were placed in an orbit such that it could reliably detect positrons with energies above a lower threshold of $5 \, {\rm MeV}$, then a few years' worth of data could either place strong constraints on PBH mass functions peaked in the as-yet unconstrained ``asteroid-mass'' window, or yield the first-ever positive detection of a PBH and its tell-tale Hawking emission.

Because such a dataset does not currently exist, we plan to extend our current work to analyze the prospects for PBH detection with data from the PAMELA cosmic ray detector, which reported positron fluxes down to 50 MeV \cite{mikhailov_spectra_2021}---an energy range accessible due to its quasi-polar orbit at $70^o$ inclination. The most significant future work, however, will include a generalization of this analysis procedure to both $\gamma$-ray and X-ray cosmic ray datasets to probe heavier, colder PBHs. We also intend to further develop the parameter estimation techniques discussed in this work in order to analyze real cosmic ray data and search for signals of PBHs in our own Solar System. 
 
Additional next steps include usingour simulation code to determine the optimal detector parameters, orbit, and particle energy ranges such that a purpose-built instrument could probe the largest range of PBH masses and therefore the largest region of model parameter space. This result could be used to inform the design of future cosmic ray experiments with the aim of detecting Hawking radiation from PBHs. 

Additionally, detecting local PBH transits affords a variety of possible multi-messenger signals, which should be leveraged to both confirm the observation of PBH candidates and to search a wider parameter space than what is accessible via only one signature. In addition to the gravitational-perturbation signatures mentioned above \cite{bellinger_solar_2023,Tinyakov:2024mcy,tran_close_2023,Cuadrat-Grzybowski:2024uph,DeLorenci:2025wbn}, one could also consider detectable $\gamma$-ray, X-ray, and neutrino fluxes from Hawking radiation. Given our focus here on positron-producing PBHs, a natural companion signature to consider would be high-energy photons at $E_\gamma \geq m_e = 511 \, {\rm keV}$. Current constraints on PBH number density from isotropic fluxes of 511 keV photons already place strong constraints on extended mass functions up to $\mathcal{O}(10^{17}) \, {\rm g}$ \cite{DeRocco:2019fjq, Laha:2019ssq}. 

We may estimate an upper bound on the possible 511 keV photon flux from the PBH transits discussed in this article. The secondary photon emission spectra for a PBH with $M = 10^{15} \, {\rm g}$---approximately the lightest and hottest PBH that could still exist today---is of the order $d^2N_{511}^{(2)}/dtdQ \sim 10^{22} \, {\rm GeV}^{-1}{\rm s}^{-1}$. This serves as an upper bound on the emitted flux at 511 keV for all PBHs with $M < 10^{15} \, {\rm g}$ because heavier PBHs produce fewer secondary particles. 

If a PBH with $M = 10^{15} \, {\rm g}$ were to transit past Earth with an impact parameter of $1 \, {\rm AU}$, the expected flux of 511 keV photons at Earth would be $\Phi_{511} = 1.8\times10^{-5} \, {\rm m}^{-2}{\rm s}^{-1}$. The reported isotropic background flux at 511 keV is $\mathcal{O}(10^{-2}) \, {\rm m}^{-2}{\rm s}^{-1}$, indicating that the signal count rate would be about 1000 times smaller than the background. However, a $\gamma$-ray experiment with good angular resolution pointed at the transiting PBH may be able to detect it as a localized point-source. Application of a matched filter analysis, such as that described in this article, to time-series photon data would be an important technique with which to extract transit signals from low-SNR data.

\begin{acknowledgements}
We are grateful to Bryce Cyr, Benjamin Lehmann, Vincent Vennin, and Rainer Weiss for helpful discussions. This material is based upon work supported by the National Science Foundation Graduate Research Fellowship under Grant No.~2141064. Portions of this work were conducted in MIT's Center for Theoretical Physics -- a Leinweber Institute and partially supported by the U.S.~Department of Energy under Contract No.~DE-SC0012567.
\end{acknowledgements}


%

\end{document}